\documentclass[final,12pt,5p]{elsarticle}

\usepackage{amssymb}
\usepackage{placeins}
\usepackage{gensymb}
\usepackage{tablefootnote}

\usepackage{float}
\usepackage{breqn}
\usepackage{booktabs}
\usepackage[english]{babel}
\usepackage{color}
\usepackage{dblfloatfix}
\usepackage{subfig}
\usepackage{threeparttable}
\usepackage{hyphenat}
\journal{Journal of Molecular Liquids}

\begin{document}

\begin{frontmatter}



\title{Effects of additives on oil displacement in nanocapillaries: a mesoscale simulation study}


\author[1]{Guilherme C. Q. da Silva}
\author[2]{Ronaldo Giro}
\author[1]{Bruno A. C. Horta}
\author[2]{Rodrigo F. Neumann}
\author[2]{Michael Engel}
\author[2]{Mathias B. Steiner}

\address[1]{Instituto de Química - Universidade Federal do Rio de Janeiro, Avenida Athos da Silveira Ramos 149 - Bloco A, Sala 609, CEP 21941-909 , Rio de Janeiro, RJ, Brazil}
\address[2]{IBM Research, Av. Pasteur 138/146, CEP 22290-240, Rio de Janeiro, RJ, Brazil}

\begin{abstract}
We performed mesoscale simulations in order to investigate the effects that additives, such as surfactants and polymers, have in the oil displacement process by water and brine injection. A Many-Body Dissipative Particle Dynamics (MDPD) model was parameterized in order to reproduce physical properties obtained either by experiments or Classical Molecular Dynamics (MD). The MDPD model was then employed to simulate the displacement of n-dodecane by water and brine, with surfactant concentrations ranging from 1 wt. \% to 10 wt. \% SDS and polymer concentration of 0.5 wt. \% HPAM. We observed that while the additives may enter the capillary without clogging, specific combinations of surfactant concentrations and injection rates may lead to poor oil displacement. This result can be understood in terms of the balance between water-wall and (surface-mediated) water-oil interactions. As a consequence, our results show an effective reversal of the wettability character, from water-wet to oil-wet, as a function of the interface speed. We conclude by summarising the implications such results have on the fluid-fluid displacements involving additives.
\end{abstract}


\end{frontmatter}


\section{Introduction}
\label{int}
The fluid displacement process has great importance in different technological areas, ranging from soil remediation, printing, microfluidics and oil recovery. 
Therefore, a vast number of research studies regarding this process (specially in porous media) have been published in the scientific literature as reviewed by Singh \textit{et al.} \cite{singh2019}.

In the technological field of enhanced oil recovery (EOR), the residual oil that remains trapped in the reservoir after the primary recovery can be extracted by different methods, which involve increasing the mobility of oil or of the displacing phase. Heat injection is used if one is interested in increasing the mobility of the oil phase, whereas the displacing fluid mobility may be altered by the chemical flood technique \cite{petroleumguide, sheng2011}. 

The chemical flood technique may be divided according to the nature of the additives used, \textit{i.e.} polymer or surfactant flooding, which operate by different mechanisms. The enhanced recovery using surfactant flooding is associated with an increase in the capillary number caused by the decrease of the interfacial tension between water and oil. As a consequence,  viscous forces overcome the capillary forces. On the other hand, polymer flooding is primarily associated with an increase in the sweep efficiency \cite{sheng2011}. The increase in the sweep efficiency is related to an increase in the viscosity of the injection fluid upon polymer addition,  leading to a reduced mobility ratio between the displacing and the displaced phase, and also reducing the viscous protrusion effect \cite{petroleumguide, sheng2011}. 

Atomistic and mesoscale simulation techniques such as molecular dynamics (MD) and dissipative particle dynamics (DPD) are of great importance in order to relate different phenomena to the microscopic behaviour of molecular systems. In the context of the fluid displacement process, a number of simulation studies have been performed \cite{Chen2014, Chen2012, Sedghi2014, Yan2016, tang2018, tang2019}. 

 Chen \textit{et al.} used a many-body dissipative particle dynamics (MDPD) model to study the spontaneous \cite{Chen2014} and forced \cite{Chen2012} displacement of generic fluids based on their previous studies of fluid drain\-age or imbibition \cite{Chen2010}. The authors derived a li\-near relationship to track the time evolution of the interface by introducing a slip length on the mo\-dified Lucas-Washburn equation as proposed by Martic \textit{et al.} \cite{Martic2002}.
 That relationship was closely followed by their MDPD simulation results for the spontaneous displacement \cite{Chen2014}. In the case of the forced displacement, 
the recovery was reduced since the interactions between the fluids became stronger than the interactions between the injection fluid and the wall  \cite{Chen2012}. 


Sedghi \textit{et al.} \cite{Sedghi2014} performed coarse-grained molecular dynamics simulations to study the influence of the pore shape on the oil displacement. They performed two computational experiments: one involving oil displacing water from nanopores with different shapes and the other involving water flooding simulations in a system with a different wettability. The authors observed that circular pores presented the lowest threshold capillary pressure.

Yan and Yuan \cite{Yan2016} carried out atomistic molecular dynamics simulations to investigate the role of surfactants in the chemical flooding process. The authors simulated an oil drop confined in a water filled nanocapillar with and without the addition of surfactant molecules. They observed that the surfactant molecules acted by disturbing the microstructure of the oil droplet and by facilitating the surface detachment and pulling process.

One of the main mechanisms used to explain the action of surfactants in spontaneous oil droplet detachment from surfaces was proposed by Kolev \textit{et al.} \cite{KOLEV2003} after performing direct microscopic observations of the process. The mechanism was said to follow three subsequent stages: ($i$) the adsorption of surfactant molecules on the oil/water interface, leading to a shrink in the contact line; ($ii$) the diffusion of surfactant and water molecules to the region underneath the oil droplet (between the oil and the solid surface); and ($iii$) the detachment of the oil droplet from the solid surface (i. e. the wedge effect) caused by the instability of the surfactant-rich oil-solid interface. 

The above described mechanism of oil droplet detachment was later verified by classical molecular dynamics simulation \cite{liu2012}. A similar mechanism was observed in MD simulations for the detachment of an oil layer above a silica wall \cite{tang2018} due to the formation of water channels or cracks in the oil layer. In the latter study, the surfactant molecules were able to disturb the oil layer structure, thus facilitating the interaction between the water and the wall molecules. Tang \textit{et al.} \cite{tang2018} suggested that the Coulombic interactions play an important role in water channel formation. In addition they highlight the importance of water-mediated hydrogen bonds during the detachment stage. The authors  also concluded that the process is hindered by a faster water flow, since this tend to make more difficult the channel penetration by water molecules.

More recently, Tang \textit{et al.} \cite{tang2019} performed MD simulations to investigate the microscopic details of oil droplet detachment under the action of water flow. The authors also suggested a similar three stage mechanism: ($i$) the deposition of surfactant micelles on the  oil/water interface; ($ii$) the migration of surfactant molecules from the micelles onto the interface; and ($iii$) the detachment itself. The water flow caused a deformation of the droplet, tilting it in the direction of the flow, and promoted the displacement of surfactants to the rear face of the droplet in relation to the flow direction. The authors called that a {\it flooding from rear} phenomenon. Finally, they  compared the efficiency of two surfactants by calculating the droplet mobility and the detachment time.

To the best of our knowledge, the major studies concerning the action of surfactant in the oil recovery process focused solely on the oil droplet detachment process. In the present article, the oil recovery process is studied from a different perspective, with the main objective of investigating the microscopic behaviour of the displacement of an oily fluid completely filling a capillar tube by the injection of a fluid consisting of water or brine, with or without additives.
For such purpose, and due to the enormous size and time scales involved, which prohibits the use of atomistic models, a many-body dissipative particle dynamics (MDPD) model was developed and applied. 

In this study, we address the challenge of using DPD simulations in the fluid-fluid displacement phenomena.
We present an overview of the MDPD method, the methodology employed to develop a suitable model to study the systems under consideration, as well as the simulation protocol designed to analyze the effects of additives on water-oil displacement. The results concerning the effects of capillary size, surfactants and polymers on the water-oil and brine-oil fluid-fluid displacement are presented and discussed.

\section{Computational Details}
\label{comp}

The LAMMPS package  \cite{plimpton1995} was used to perform both many-body dissipative particle dynamics (MDPD) and classical molecular dynamics (MD) simulations. The MD simulations were carried out to extract a set of properties used in the obtention of the parameters needed for the MDPD simulations. These MDPD simulations were performed to investigate the oil displacement process in a capillar vessel of nanometric dimensions. This section is organized in three subsections: ($i$) a brief overview of the MDPD method; ($ii$) the strategy for obtaining the MDPD parameters used in the present work; ($iii$) the computational flow experiment designed to study the influence of additives on the oil displacement process.

\subsection{Overview of the MDPD method}
The dissipative particle dynamics (DPD) me\-thod, originally developed by Hoogerbrugge and Koelman \cite{Hoogerbrugge_1992}, is suitable to simulate mesoscale systems since it can reach longer timescales and treat larger size scales compared to atomistic simulations. This is due to the coarse graining of the degrees of freedom and to the possibility of using larger integration time steps due to the soft nature of the potentials.

The original DPD model was not able to describe the vapor-liquid equilibrium of pure substances. Because of that, the MDPD \cite{Pagonabarraga2001, Warren2003, Trofimov2005} method was introduced in order to reproduce the correct thermodynamic behaviour of interacting systems. In this work, the MDPD formalism proposed by Warren \cite{Warren2003} was used and its equations for the interaction forces are given by:

\begin{equation}
\label{mdpd}
\vec{F}_{ij} = \vec{F}_{ij}^{C} + \vec{F}_{ij}^{D} + \vec{F}_{ij}^{R} \quad ,
\end{equation}

\noindent where $\vec{F}_{ij}^{D}$ and $\vec{F}_{ij}^{R}$ are, respectively, the dissipative and random forces, responsible to model the thermostat acting on the system. Usually, these forces take the form proposed by Espa\~nol and Warren \cite{pepe1995}: 

\begin{equation}
\label{dissipative}
\vec{F}_{ij}^{D} = -\gamma_{ij} \omega_{D}(\vec{r}_{ij}) (\vec{v}_{ij} \cdot \vec{e_{ij}})  \vec{e_{ij}}
\end{equation}

\begin{equation}
\label{random}
\vec{F}_{ij}^{R} = \sigma \omega_{R} (\vec{r}_{ij}) \zeta_{ij}   \quad ,
\end{equation}

\begin{equation}
\label{wrwd}
\omega_{D} (r_{ij})  = \omega_{R} (r_{ij})^{2} \quad ,
\end{equation}

\begin{equation}
\label{gamma-sigma}
\sigma^{2} = 2 \gamma_{ij} k_B T \quad ,
\end{equation}

\noindent where ${r_{ij} = | \vec{r}_i - \vec{r}_j |}$ and $\vec{e}_{ij} = (\vec{r}_i - \vec{r}_j)/{r_{ij}}$.

In Equation \ref{dissipative} , $\vec{v_{ij}} = \vec{v}_{i} -\vec{v}_{j}$, where $\vec{v}_i$ is the velocity of the ith particle, $\gamma_{ij}$ is the friction coefficient and $\omega_{D}$ is a weight function responsible to address the range of action of this force. Equation \ref{random} models the random forces, where  $\omega_{R}$ is a weight function, $\zeta_{ij}$ is a gaussian white-noise with special stochastic properties \cite{pepe1995} and with amplitude $\sigma$. Note that to ensure the fluctuation-dissipation theorem, the respective weight functions, noise amplitude and friction coefficient are related according to Relations \ref{wrwd} and \ref{gamma-sigma} \cite{pepe1995}, in which $k_B$ is the Boltzmann constant and $T$ the absolute temperature. For more details, the reader is referred to the fundamental papers of Espa\~nol and Warren \cite{pepe1995, Warren2003}.

The remaining term in Equation \ref{mdpd} is the conservative force and can be expressed as follows:

\begin{dmath}
\label{conservative}
F_{ij}^{C} = B_{ij}\left( \bar{\rho_i}+\bar{\rho_j}\right)\omega_{C,r}(r_{ij}) - A_{ij}\omega_{C,a}(r_{ij})+F_{iof} \quad 
\end{dmath}

\begin{equation}
\label{omegap}
 {\bar{\rho_i}} = \sum_{j \ne i} \omega_p (r_{ij}) \quad ,
\end{equation}

\begin{equation}
\label{omegapder}
-\omega_p'(r_{ij}) = \frac{\omega_{C,r}(r_{ij})}{2\alpha} \quad ,
\end{equation}

\begin{equation}
\label{alfamf}
\alpha = \frac{2\pi}{3} \int_{0}^{\infty} r^3 \omega_{C,r}(r) dr
\end{equation}

According to Warren \cite{Warren2003}, the conservative force is represented as a sum of a repulsive and an attractive contribution, which are the first and second terms, respectively, in Equation \ref{conservative}. In addition, the last part of this Equation accounts for the force concerning the internal degrees of freedom such as bond stretching and angle bending.

The second term in Equation \ref{conservative} is linear concerning the distance between the particles and the first represents a functional dependence on the local density of particles (Equation \ref{conservative}) \cite{Warren2003}. In this relation, $A_{ij}$ and $B_{ij}$ are, respectively, the attractive and repulsive force amplitudes; $\omega_{C,a}$ and $\omega_{C,r}$ are the respective weight functions.

${\bar{\rho_i}}$ is a weighted local density function which depends on the distances between the neighbor particles as it is shown by Equation \ref{omegap}. The form of this function, that is $\omega_p (r_{ij})$, is not arbitrary. As was shown by Warren \cite{Warren2003}, the existence of a potential function, which the gradient is related to the forces acting on the particles, can be satisfied if the first derivative in relation to distances of $\omega_p (r_{ij})$ is proportional to the repulsive force weight function $\omega_{C,r}$ (Equation \ref{omegapder}). Moreover, the proportionality constant ($\alpha$) will also depend on the functional form of
$\omega_{C,r}$. Combining the virial equation of state and the mean field theory, Warren wrote this relation in the form of Equation \ref{alfamf} \cite{Warren2003}.

The weight functions are chosen to vanish beyond a cutoff distance $r_c$ and are usually linear with the form given by Equation \ref{weight} \cite{Warren2003}, but this is not mandatory. In the present article, another form of the weight function for the repulsive force was used. The $\omega_{C,r}$ was chosen to be a Lucy function (Equation \ref{Lucy}), which is already implemented in the LAMMPS package \cite{Lisal2011, Brennan2014, Moore2016} and was used in other MDPD studies \cite{Tiwari2006, Yamada2015} and other mesoscale models \cite{Moore2016}.

\begin{equation}
\label{weight}
\omega_{C,a}\left(r_{ij}\right) = \left(1-\frac {r_{ij}} {r_c}\right) 
\end{equation}

\begin{equation}
 \omega_{C,a}\left(r_{ij}\right) =\omega_{R}\left(r_{ij}\right) =\omega_{D}\left(r_{ij}\right) 
\end{equation}

\begin{equation}
\label{Lucy}
\omega_{C,r}\left(r_{ij}\right) =  \left( 1 + \frac{3r_{ij}}{r_D}\right)\left(1-\frac{r_{ij}}{r_D}\right)^3
\end{equation}

It is important to note that MDPD versions that explicitly account for the electrostatic interactions have emerged \cite{Mao2015, Mayoral2016}. However it has been shown \cite{Ghoufi2012} that the conventional MDPD may be used to model charged systems as well. Thus, in this work a conventional MDPD model coupled with a Lucy weight function  for the repulsive forces was used.

\subsection{MDPD Parameters}

Several strategies have been reported in the literature for obtaining  parameters of DPD models. Some of them rely on parameter fitting in order to reproduce relevant experimental quantities as in the case of the first models using the DPD method \cite{Groot1997}, where the interaction parameter was fitted so as to reproduce the fluid compressibility. Other property that is typically used for the obtention of the cross-interaction parameter is the interfacial tension between two fluids \cite{Maiti2004}.

Groot and Warren (1997) \cite{Groot1997} suggested that the DPD method could be regarded as a continuous version of a lattice model and compared the equations with the Flory-Huggins model. They were able to establish a relationship between the DPD cross-interaction parameters and the Flory-Huggins interaction parameters. This strategy was further generalized  for the MDPD method in the work of Jamali \textit{et al.} \cite{Jamali2015}.

The relationship between the Flory-Huggins and the Hildebrand solubility parameters can be explored to obtain the DPD cross-interaction parameters \cite{Maiti2004, Mayoral2016}. However, these parameters are not defined for charged systems and, in such cases, it is common to relate the DPD parameter to the pair contact energy \cite{Ryjkina2002, Ghoufi2012, Ghoufi2013}. It is important to mention other strategies to derive the DPD interaction parameters that involve the exploration of the excess free energy models \cite{ALASIRI2017}, water-octanol partition coefficients \cite{anderson2017} and more elaborate statistical approaches \cite{canchaya2016}.

In the present study, the necessary parameters were obtained in order to reproduce critical experimental quantities that are related to the  phenomena under investigation. When this strategy was impossible or inconvenient, the parameters were obtained to reproduce structural quantities obtained by classical molecular dynamics simulations using the CHARMM force field \cite{charmm1} and a CHARMM-compatible parameter set for the silica walls \cite{cruz2006, lorenz2008}. All the fitting procedures performed in this work were carried out in a systematic trial and error basis. 

All MD simulations were performed at the NPT ensemble with the Nos{\'e}-Hoover thermostat and the MTK barostat \cite{mtk} at 300 K and 1 atm with the damping parameters of 100 fs and 1000 fs respectively. After equilibration was observed, at least 1 ns of production was carried out for the necessary sampling. Except for the simulations with the silica wall, periodic boundary conditions were used. Both MD and MDPD simulations boxes were built with the aid of the Packmol code \cite{pack} and the VMD software \cite{HUMP96}.

\begin{figure}[h]
\centering
\includegraphics[width=0.7\columnwidth]{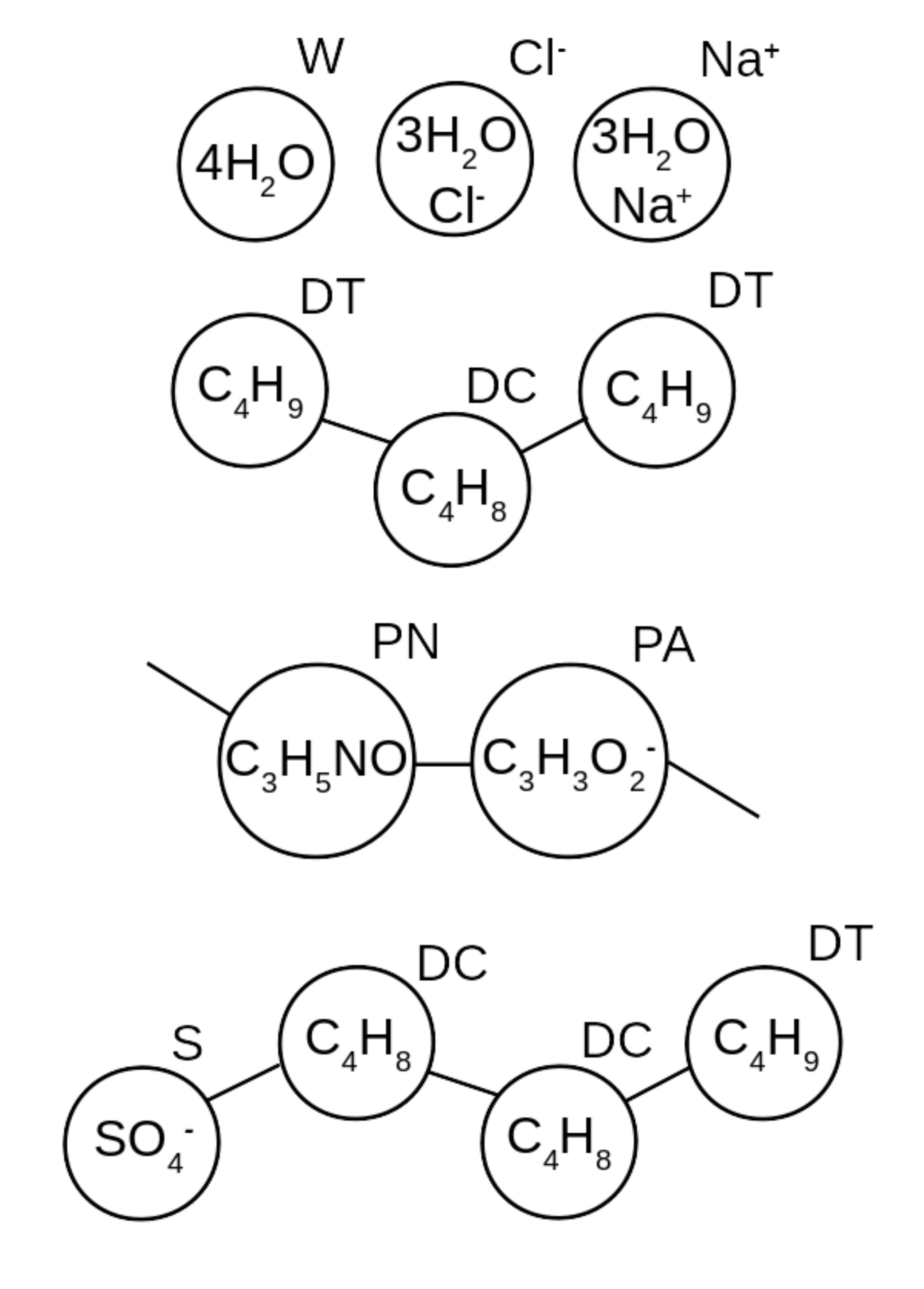}
\caption{Coarsing level of molecules and bead types nomenclature}
\label{beads}
\end{figure}

The systems of interest consisted of a silica capillary tube filled with n-dodecane subjected to the displacement by water or brine containing HPAM or SDS in different concentrations. Figure \ref{beads} displays the coarsing level of the molecular and supramolecular models, along with the naming scheme for the bead types. Since the water model dictated the modeling of the other components, its parametrization is first discussed followed by the methodology employed to obtain the other parameters. Table \ref{scheme1} lists the model parameters along with the properties utilized as target in the calibration procedure.

{\small
\begin{table}[]
\setlength{\tabcolsep}{4pt}
\caption{\label{scheme1} Summary of the relevant parameters and the model or property utilized to obtain it.}
\begin{tabular}{cl}
\hline
Parameter                                                    & Model or target property                                                                             \\ \hline
$A_{w-w}$                                                    & Scaling of Ghoufi and Malfreyt model \cite{Ghoufi2012}                                                                  \\
$B$                                                          & Scaling of Ghoufi and Malfreyt model \cite{Ghoufi2012}                                                                   \\
$r_D$                                                        & PV diagram                                                                   \\
$\gamma_w$                                                   & Viscosity                                                                    \\
$A_{DC-DT}$                                                  & Density                                                                   \\
$\gamma_D$                                                   & Viscosity                                                                 \\
$A_{W-Na^{+}}$                        & Density                                                                \\
$A_{W-Cl^{-}}$                         & Density                                                                 \\
$A_{Na^{+}-Cl^{-}}$ & Density                                                                 \\
$A_{W-DC}$                                                   & Interfacial tension                                                      \\
$A_{W-DT}$                                                   & Interfacial tension                                                  \\
$A_{Na^{+}-DC}$                                              & Set up a moderate attractive interaction                                     \\
$A_{Na^{+}-DT}$                                              & Set up a moderate attractive interaction                                     \\
$A_{Cl^{-}-DC}$                                              & Set up a moderate attractive interaction                                     \\
$A_{Cl^{-}-DT}$                                              & Set up a moderate attractive interaction                                     \\
$A_{S-W}$                                                    & Interfacial tension near the CMC                         \\
$A_{S-DC}$                                                   & {[}SDS{]} x interfacial tension curve                        \\
$A_{S-DT}$                                                   & {[}SDS{]} x interfacial tension curve                        \\
$A_{S-S}$                                                    & {[}SDS{]} x interfacial tension curve                        \\
$A_{S-Na^{+}}$                                               & {[}SDS{]} x interfacial tension curve                        \\
$A_{S-Cl^{-}}$                                               & Interfacial tension near the CMC \\
$A_{PN-W}$                                                   & Radius of gyration                                             \\
$A_{PN-DC}$                                                  & Radius of gyration                                          \\
$A_{PN-DT}$                                                  & Radius of gyration                                          \\
$A_{PA-W}$                                                   & Radius of gyration                            \\
$A_{PA-DC}$                                                  & Radius of gyration                         \\
$A_{PA-DT}$                                                  & Radius of gyration                         \\
$A_{PA-PA}$                                                  & Radius of gyration                            \\
$A_{PN-PA}$                                                  & Radius of gyration                        \\
$A_{PA-PA}$                                                  & Radius of gyration                        \\
$A_{PN-Na^{+}}$                                              & Radius of gyration                                                               \\
$A_{PN-Cl^{-}}$                                              & Pair correlation function                                                               \\
$A_{PA-Na^{+}}$                                              & Radius of gyration                                                              \\
$A_{PA-Cl^{-}}$                                              & Pair correlation function                                                               \\
$A_{Si-Si}$                                                  & Model of Henrich \textit{et al.}\cite{Henrich2007}                                                                                \\
$A_{Si-W}$                                                   & Contact angle                                                          \\
$A_{Si-DC}$                                                  & Contact angle                                                       \\
$A_{Si-DT}$                                                  & Contact angle                                                       \\
$A_{Si-S}$                                                   & Pair correlation function                                                               \\
$A_{Si-Na^{+}}$                                              & Pair correlation function                                                               \\
$A_{Si-Cl^{-}}$                                              & Pair correlation function                                                               \\
$k_b$                                                        & Iterative Boltzmann Inversion method                                                                   \\
$k_\theta$                                                   & Iterative Boltzmann Inversion method                                                                   \\
$b_0$                                                        & Iterative Boltzmann Inversion method                                                                   \\
$\theta_0$                                                   & Iterative Boltzmann Inversion method                                                                   \\ \hline
\end{tabular}
\end{table}
}

The water MDPD model of Ghoufi and Malfreyt \cite{Ghoufi2012}, which was rescaled from a supramolecular coarse grain level of 3 to 4, was used. The cutoff radius for truncating the conservative force  ($r_C$) was adjusted in order to maintain the dimensionless density reported in the work of Ghoufi and Malfreyt, which led the dimensional density to remain unchanged. 

According to Groot and Rabone \cite{rabone2001}, both the attractive ($A$) and repulsive ($B$) force amplitudes should be linearly scaled when the coarse grain level is modified. Note, also, that the No-go theorem \cite{Warren2013} implies that $B$ is universal ($B_{ij}=B$ for all pairs $ij$) so the system behaves as Hamiltonian.

The repulsive force cutoff radius ($r_D$) had to be adjusted due to the use of a different weight function. This parameter was fitted to reproduce the water pressure-volume diagram at 300 K. 

The water friction coefficient ($\gamma_w$) was obtained by fitting the Gaussian noise amplitude ($\sigma_w$) to reproduce the water viscosity that was calculated using the reverse Poiseuille flow methodology \cite{Fedosov2010}.

With the chosen coarse-grained level, the n-dodecane molecules consisted of 3 beads. The middle bead was attributed a different type than the other beads. The interaction parameter between the center bead and the two terminal beads was fitted to reproduce the n-dodecane density at 300 K. The n-dodecane friction coefficient ($\gamma_d$) was obtained using the same procedure discussed for the water case.

The dissipative force was calculated in such a way that any particle (except the beads of the n-dodecane molecule) interacting with water experimented a dissipative force proportional to $\gamma_w$. All other dissipative forces were considered proportional to $\gamma_d$, according to the recommendations of Novick and Coveney \cite{novik2000}.

The brine solution model consisted of three types of bead: a water, a sodium and a chloride bead. Each ion bead consisted of a supramolecular model of one ion solvated by three water molecules. The initial parameters were also taken from  Ghoufi and Malfreyt \cite{Ghoufi2012}, but all cross-inter\-action parameters were simultaneously scaled to reproduce the experimental density of a 5.1 M sodium chloride aqueous solution.

The water-dodecane interactions were adjusted to reproduce the experimental interfacial tension reported by Zeppiere \textit{et al.} \cite{Zeppieri2001} using the Irving-Kirkwood formulation \cite{kirkwood1950} and the same simulation setup reported by Ghoufi and Malfreyt \cite{Ghoufi2010, Ghoufi2013, Ghoufi2016, Ghoufi2011}. Note that different box sizes were tested and not only the NPT ensemble was used but also the NPzzT, that is, only the stress tensor component normal to the interface was coupled to the barostat. Since none of those variations resulted in appreciable difference in the values for the water-dodecane interfacial tension, it is expected that size and methodological effects did not have much influence on the parameterization.

The interaction between ions and the apolar phase is probably not so important for the system behaviour and a less thorough procedure was used to derive the corresponding parameters. These parameters were set to model an interaction less attractive than the one between water and the ions but more attractive than the interactions between water and n-dodecane.

The interactions involving the sulfate, water and sodium beads, were adjusted to reproduce the expected behaviour of the water-hydrocarbon interfacial tension decrease with the increase in the surfactant concentration. The sulfate-water interaction was adjusted to reproduce the experimental interfacial tension of the system in the vicinity of the critical micellar concentration \cite{Oh1993}. All other parameters were simultaneously adjusted to approximate the simulated interfacial tension decrease curve to experimental results taken from the work of Rehfeld \cite{Rehfeld1967} for water-SDS-decane and water-SDS-heptadecane interfaces. Note that the experimental results were later corroborated by other papers in the literature \cite{ Joos1990, Oh1993, Cortes-Estrada2014}.

\begin{figure}[ht]
\centering
\includegraphics[width=0.5\textwidth]{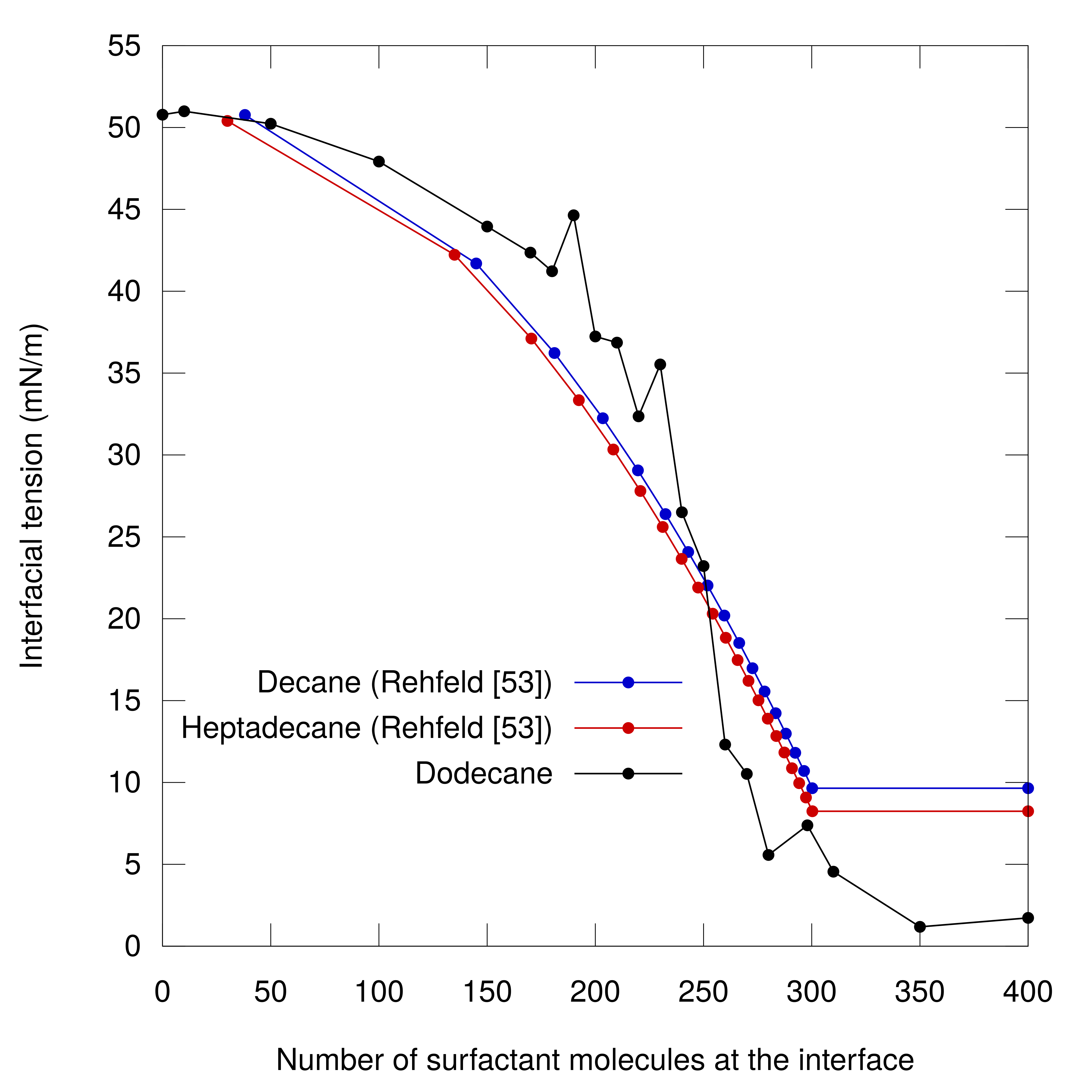}
\caption{Interfacial tension as a function of the number of surfactant molecules at the interface. For comparison with the simulation results that overestimates the CMC, the experimental curve was prolonged to a fictitious number of surfactants at the interface (above the number of surfactant molecules at the CMC).
This clearly shows that the model only reaches the CMC after about 350 molecules saturate the interface.}
\label{st}
\end{figure}

Figure \ref{st} displays the decrease of the interfacial tension for the simulated water-SDS-n-dodecane interface and compares it to the experimental results of Rehfeld \cite{Rehfeld1967} for different hydrocarbon interfaces. Note that, in order to perform this comparison, the adsorption model derived by Rehfeld \cite{Rehfeld1967} was used to calculate the number of surfactant molecules at an interface with the same area as the one used in our simulations (blue and red lines in Figure \ref{st}). 
The simulated values are reported considering that all surfactant molecules were at the interface. 
The model is in good agreement until the vicinity of the critical micellar concentration but instead of reaching a low level pla\-teau, the interfacial tension continues to decrease, indicating that the model overestimates the number of surfactants at the interface. In other words, this suggests that the CMC of the model is slightly above the experimental one. Each point in the curve refers to a 20 ns simulation. 


The chlorine-sulfate interaction parameter was  adjusted to account for the reported pronounced decrease in the interfacial tension for the brine-SDS-dodecane system \cite{kumar2016}. All the interfacial tension calculations followed the same protocol as mentioned before for the parameterization of the water-dodecane interactions.

Two simulation movies are made available in the supplementary material displaying the comparison between the interfaces of water-SDS-n-dodecane and brine-SDS-n-dodecane. One can see that the main difference is that the presence of the salt ions hold more surfactant molecules at the interface by making the micelle formation more difficult. The consequence of this is a more pronounced decrease in the interfacial.

Concerning HPAM, it is reported, in the literature \cite{rabiee2005}, that its hydrolysis degree  impacts its gyration radius $R_g$ and consequently its viscosity, since both quantities are related \cite{Kok1981}. The hydrolysis of the acrylamide units into acrylate will stretch the chain due to the repulsion between the charged moieties. Because of that, the MDPD parameters were obtained in order to reproduce this behaviour following the procedure discussed below.

The model was defined as a chain of 30 connected beads representing acrylate or acrylamide units. The parameters related to the interactions between acrylamide with water and with dodecane were fitted to reproduce $R_g$ obtained by classical molecular dynamics of the non-hydrolyzed  acrylamide polymer in either aqueous or dodecane solution.
The same procedure was repeated for the hydrolyzed polymer considering a 100 \% hydrolysis ratio in order to describe its interactions with water and dodecane. The parameter concerning the interaction between the acrylamide and acrylate units was fitted to reproduce $R_g$ obtained from a MD simulation  of  a polymer with a 25 \% hydrolysis  ratio. The parameters regarding the interactions of the two ion beads with the polymer units were adjusted in different manners. Since the interaction between the cation bead with the polymer beads are more relevant, the attractive parameter for the interaction between the PN and PA beads with the Na$^+$ beads were adjusted so as to reproduce the gyration radius of the non-hydrolyzed and the fully hydrolyzed polymer in brine, respectively. This was done by fitting to the radius of gyration obtained with the MD simulations. Finally, the interaction parameter between the polymer beads and the anion beads were fitted to reproduce the corresponding radial distribution functions ($g(r)$) obtained via MD.

In order to model the silica wall, the parameters derived by  Henrich \textit{et al.} \cite{Henrich2007} were used. This model treats the wall as a phase consisting of randomly distributed particles with restrained mobility. The parameters corresponding to the interaction between the two fluids and the wall were fitted to reproduce the contact angle obtained by MD simulations. Simulations consisting of a half-filled nanocapillary with a diameter of 10 nm were carried out and the curvature of the fluid-vapor interface was used to calculate the static contact angle. All other interaction parameters involving the silica wall were fitted to reproduce the corresponding  pair correlation functions obtained by MD simulations. 

Intramolecular interactions of all species presented in the system, \textit{i.e.} the bond stretching and angle bending potentials, were obtained using the Iterative Boltzmann Inversion (IBI) method \cite{Bayramoglu2012, Agrawal2014}. For that, it was necessary to perform MD simulations consisting of pure dodecane; an SDS aqueous solution and the polymeric aqueous solution at 1 atm and 300 K. In these simulations, the distribution of the internal degrees of freedom were sampled and used to perform the IBI analysis. More precisely, after obtaining the potential form, an harmonic fitting was performed to estimate the force constants and the reference bond distances and angles.

The final MDPD model was then tested based on a simulation of water-n-dodecane system inside a 10 nm diameter silica capillary tube. The calculated contact angle of 69\degree{}  agrees well with previously reported results  \cite{BI2004, jung2016}.

The simulation parameters are listed in  Tables \ref{radii}-\ref{mdpd_parms} and a comparison between the calculated properties obtained with the MDPD model and the corresponding reference values used in the fitting procedure are made available in the Supplementary Material. Table \ref{radii} lists the cutoff radii, repulsive force constant and time step. Table \ref{friction} presents the Gaussian noise amplitudes related to the friction coefficients. Table \ref{idofs} lists the parameters concerning the bond-stretching and angle-bending potentials and Table \ref{mdpd_parms} reports the attractive force constants. Note that the latter has no entry for the interactions between  surfactant and polymer beads, since no simulated system considers them simultaneously.

\begin{table}[ht]
\centering
\caption{\label{radii} Cutoff radii, repulsive force amplitude and time step.}
\begin{tabular}{cccc}
\hline
 $r_c$ (\AA)  & $r_D$ (\AA) & $B$ (eV$\cdot$\AA\textsuperscript{2}) & $\Delta t$ (ps) \\ \hline
 9.38 & 9.06 & 62.55  & 0.01   \\ \hline
\end{tabular}
\vspace{-4mm}
\end{table}

\begin{table}[H]
\centering
\caption{\label{friction} Gaussian noise amplitude (eV$\cdot$\AA\textsuperscript{-1}$\cdot$ps\textsuperscript{-1/2})}
\begin{tabular}{cc}
\hline
 Water   & n-Dodecane \\ \hline
 0.09911 & 0.03000    \\ \hline
\end{tabular}
\end{table}

Usually, in the DPD models, all the interactions involving beads of the same type are taken to have the same conservative force amplitude ($A_{ii}$). However, this is only required so that the relationship between the DPD and the Flory-Huggins model is conserved. In this work, it was necessary to tune this parameter to increase the repulsion between the sulfate and acrylate beads. Therefore, these beads experience just repulsive forces. This issue was not observed for the sodium beads of the brine, possibly because the model accounts for the fact that water molecules shield the ion-ion interactions.

\begin{table}[H]
\caption{\label{idofs} Bond stretching and angle bending potential parameters}
\begin{tabular}{ccc}
\hline
Bond, Angle                             & $k_{B,\theta}$ (eV$\cdot$\AA\textsuperscript{-2},rad\textsuperscript{-2} )              & $r_0, \theta_0$ (\AA,\degree)              \\ \hline
DT-DC                        & 0.1503                     & 4.50                      \\
PN-PN                        & 0.7480                     & 3.96                      \\
PN-PA                        & 0.6512                     & 3.87                      \\
PA-PA                        & 0.7803                     & 3.87                      \\
S-DC                         & 1.5000                     & 4.34                      \\
DC-DC                        & 1.4000                     & 5.05                      \\
DC-DT                        & 0.1503                     & 4.50                      \\
DT-DC-DT                     & 0.0701                     & 130.0                     \\
PN-PN-PN & 1.2960 & 91.4  \\
PN-PN-PA & 0.7414 & 97.4  \\
PN-PA-PN & 0.9369 & 89.0  \\
PN-PA-PA & 0.9379 & 106.9 \\
PA-PN-PA & 0.7633 & 92.5  \\
PA-PA-PA & 1.0468 & 95.3  \\
S-DC-DC  & 0.0600 & 140.0 \\
DC-DC-DT                     & 0.0500                     & 140.0                     \\ \hline
\end{tabular}
\end{table}

\begin{table*}[b]
\caption{\label{mdpd_parms} Conservative Force amplitudes (eV$\cdot$\AA\textsuperscript{-1})}
\resizebox{\textwidth}{!}{%
\begin{tabular}{cccccccccc}
\hline
    & W       & DT      & DC      & Na      & Cl      & S       & PN      & PA      & SiO    \\ \hline
W   & -0.1517 &         &         &         &         &         &         &         &        \\
DT  & -0.1070 & -0.1517 &         &         &         &         &         &         &        \\
DC  & -0.1070 & -0.0850  & -0.1517 &         &         &         &         &         &        \\
Na  & -0.1892 & -0.1300 & -0.1300 & -0.1517 &         &         &         &         &        \\
Cl  & -0.1894 & -0.1200 & -0.1200 & -0.1675 & -0.1517 &         &         &         &        \\
S   & -0.4500 & -0.1300 & -0.1300 & -0.8000 & \phantom{-}0.0000  & \phantom{-}0.1250  &         &         &        \\
PN  & -0.1490 & -0.1300 & -0.1300 & -0.1000 & -0.1600 &         & -0.1517 &         &        \\
PA  & -0.1700 & -0.1499 & -0.1499 & -0.3000 & -0.1600 &         & -0.1600 & \phantom{-}0.1500  & -0.1800\\
SiO & -0.1460 & -0.1250 & -0.1250 & -0.0800 & -0.0800 & -0.3800 & -0.1500 & -0.1800 & \phantom{-}0.0000 \\ \hline
\end{tabular}%
}
\end{table*}

\subsection{Fluid displacement simulation}

When analyzing the fluid displacements in capillary tubes, it is important to follow the curvature dynamics of the interface. The molecular kinetic theory of Blake \cite{blake1969} (MKT) is one of the \textit{Ans\"{a}tze} used to perform such study. It assumes that the advance of the interface is an activated process. Under this assumption,  the Eyring \cite{Eyring1941} formalism, originally developed to relate the reaction rate with the energy barrier related to the process, can be used to express the fluid velocity as a function of the dynamic contact angle (Equation \ref{Blake1}) \cite{blake1969}: 

\begin{dmath}
\label{Blake1}
U=2K\lambda \times \\ sinh \left[\left(\frac{\gamma_{12}}{\Delta n k_B T}\right) \left(cos \theta_D- cos\theta_{0}\right)\right]
\end{dmath}

\noindent where $U$ is the interface velocity, $K$ is the characteristic frequency for the molecular displacements at the three-phase contact line, $\lambda$ is the distance between the sites where liquid molecules adsorb at the surface, $\gamma_{12}$ is the interfacial tension between the advancing and receding fluid, $\Delta n$ is the number of affected adsorption sites,  $k_B$ is the Boltzmann constant, $T$ is the absolute temperature, $\theta_D$ and $\theta_{0}$ are, respectively, the dynamic and static contact angles.

 All simulated systems consisted of a capillary tube of nanometric dimensions composed of silica walls and filled with n-dodecane. A reservoir with dimensions of 330x330x120 \AA{ } containing n-dodecane with a density of 0.745  g/cm\textsuperscript{3} was connected to the exit of the channel and a reservoir with dimensions of 330x330x495 \AA{ } containing the  injection fluid was connected to the entrance of the channel (Figure \ref{system}). The injection fluid consisted of water or brine in their respective densities at 300 K and 1 atm with or without surfactants and polymers. The size of both reservoirs was independent of the capillary radius. 

Each reservoir was exposed to reflective walls in 5 of its faces except the one that connected the capillary. Note that, for the oil reservoir, the face opposed to the channel exit was placed further apart in order to let the oil inside the capillary to be displaced. The face connecting the reservoirs to the channel was a silica surface wall of the same surface area of the reservoir (330x330 \AA). At the center of this silica wall, a circular hole corresponding to the channel entrance with the same radius as the capillary channel was created.

The injection fluids consisted of pure water; SDS aqueous solutions with surfactant concentrations of 1.0 and 10 wt. \%; brine with the same concentrations of surfactants and  0.5 wt. \% aqueous or brine solutions of a 20 \% hydrolized HPAM with a chain of 200 monomers. Note that the brine fluid consisted of an aqueous solution of sodium chloride with a concentration of ca. 3.5 wt. \%. 

The spontaneous and forced displacement experiments of n-dodecane by pure water were studied in capillaries with diameters of 10, 20 and 30 nm. The experiments considering the other injection fluids were carried out considering only the channel with a diameter of 20 nm. 

To model the forced displacement, the injection fluid reservoir was subjected to the action of a virtual piston, that is, the reflective wall that was placed on the face opposing the capillary entrance was set to move with a fixed velocity of 4 and 8 \AA/ns. 

Two different setups were used to simulate the surfactant effects on fluid-fluid displacement. The first approach consisted of randomly distributing the surfactant molecules  in the injection fluid reservoir according to the aforementioned concentrations. The second setup consisted of adding a surfactant layer between the water and n-dodecane phases in the capillary vessel  (See Figure \ref{system}). The number of SDS molecules in this layer was computed with the aid of the interfacial adsorption isotherms fitted by Rehfeld \cite{Rehfeld1967} for the water-SDS-decane and water-SDS-heptadecane inter\-faces.

During the displacement simulations the density profile across the capillary length was monitored. The interface progress was also tracked by means of an algorithm, which divided the simulation box in bins of 5 \AA{ } in the non-axial directions and searched for the outermost water molecules, that is the particles which traveled the most in the axial direction of the cylinder, in each bin a few molecular layers apart the wall. 

With the instantaneous interface location, the instantaneous contact angle was calculated using a fitting sphere procedure for the interface curvature. By monitoring both the interface progress and the curvature, it was possible to observe the relation between these quantities and compare with the MKT of Blake and Haynes \cite{blake1969}.

Usually, the relevant parameters of the MKT (Equation \ref{Blake1}) are treated as adjustable parameters \cite{BLAKE20061} and are fitted to reproduce experimental measurements of the dynamic contact angle evolution \cite{blake1969, Coninck2008}. Previous studies extracted  the parameters of Equation \ref{Blake1} from MD simulations by calculating the instantaneous contact angle and interface velocities \cite{Blake1997, ruijter1999, Martic2002, Martic2004, Coninck2008, Stukan2010, Ahadian2010}. However, this is only possible when the initial configuration is not very far from the equilibrium \cite{Coninck2008}, which was not the case observed in the simulations presented in this work, since the initial positions were randomly initialized and the initial velocities sampled according to a Maxwell-Boltzmann distribution. Because of that, the comparison made in this paper with the MKT is only qualitative.

\section{Results and discussion}

\subsection{Capillary tube size effects}

In order to have a suitable model to study the effects of additives on the oil displacement, simulations of the n-dodecane displacement by pure water were first performed. The simulations were carried out in cylindrical capillaries, composed of silica beads, with different diameters, namely 10, 20 and 30 nm. The main interest with this investigation was to have an estimate of a minimum diameter size at which the acting forces would no longer impede the fluid displacement.

Interestingly, the spontaneous displacement was not observed in the 10 nm diameter tube. It is known by both experiments \cite{GRANICK1374} and simulations \cite{Cui2003} that n-dodecane can exhibit an extremely high viscosity when confined. The simulations involving the 10 nm diameter tube evidenced this behaviour since the capillary forces were unable to overcome the viscous forces, holding the n-dodecane molecules together.

\begin{figure}[htb]
\centering
\includegraphics[width=0.5\textwidth]{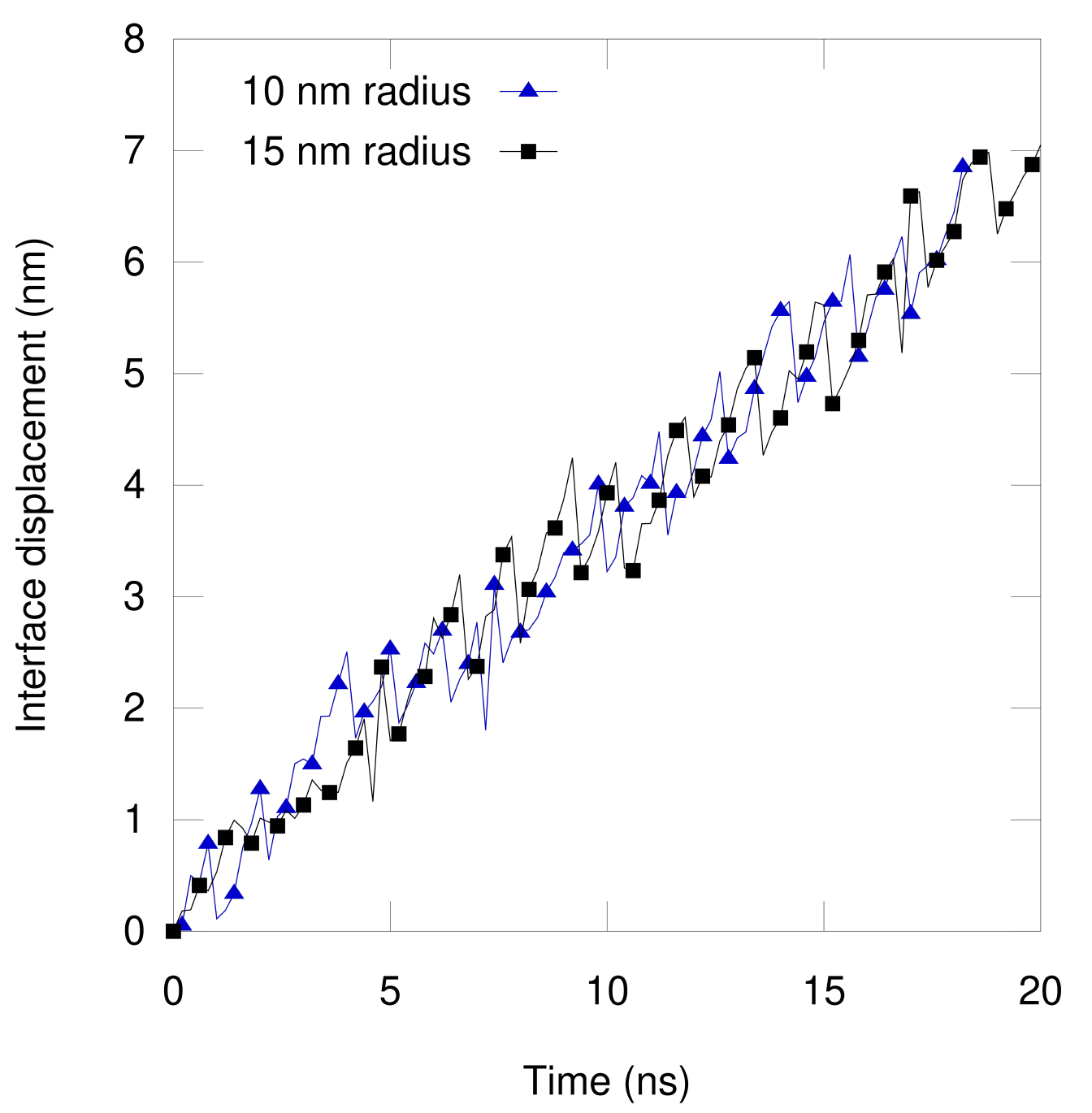}
\caption{Interface evolution in the spontaneous fluid-fluid displacement n capillaries with different radii.}
\label{radii_plot}
\end{figure}

No differences were observed in the fluid-fluid spontaneous displacement in the capillaries of 20 and 30 nm diameter (See Figure \ref{radii_plot}). Both systems presented the same interface velocity and the same average contact angle of 76\degree. Chen \textit{et al.} \cite{Chen2014} also performed MDPD simulations of the spontaneous capillary displacement and reported that the interface velocity increased with the capillary radius. However, this velocity increase was only noticeable for radial differences larger than 5 nm. This result shows that the relation between surface and bulk forces may not change considerably with this variation in the capillary diameter. Concerning an EOR scenario, these results suggest that it may be very difficult to promote fluid-fluid displacement in capillaries with diameters lower than 20 nm.

Since no significant differences were observed when comparing the 20 and the 30 nm diameter tubes, the effects of additives were studied only with the thinner tube. This reduced significantly the computational cost, since the thinner tube involved  $\sim$1.3 million particles compared to $\sim$3.65 million particles for the thicker tube.

\subsection{Surfactant effects}

The discussion concerning the surfactant effects considers mainly the results using the second approach for building the initial configuration, that is, with some surfactant mo\-le\-cu\-les placed at the interface of the fluids. This was deemed necessary, because the diffusion of surfactant mo\-le\-cu\-les to the interface could not be observed with the other approach. When the surfactant mo\-le\-cu\-les were randomly placed within the water, they started to form micelles as soon as the simulations started. Despite of that, in both setups, no surfactant adsorption at the silica wall was observed.

\begin{figure}[htb]
\centering
\includegraphics[width=0.5\textwidth]{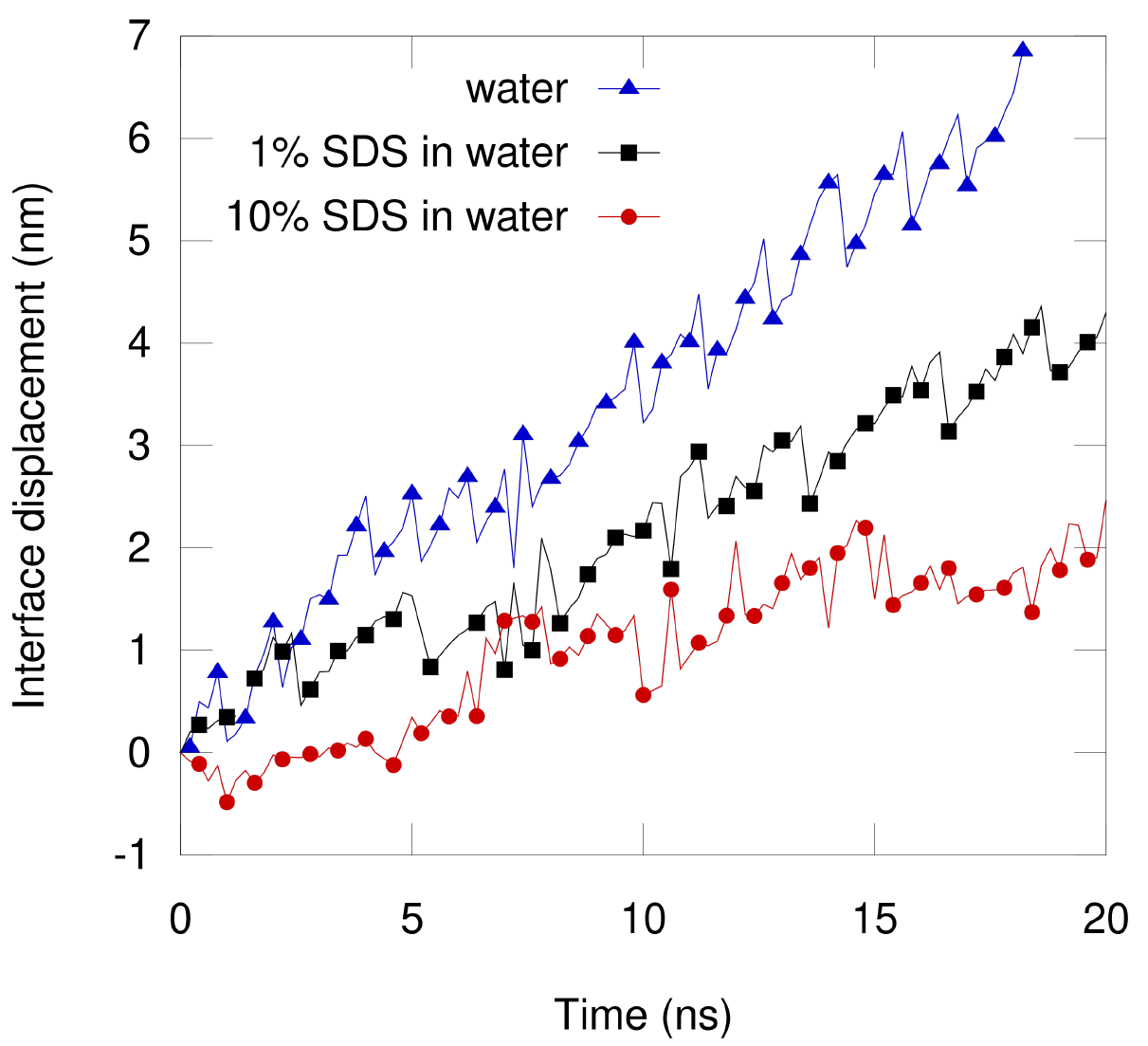}
\caption{Interface evolution in the spontaneous fluid-fluid displacement for the water-SDS-dodecane systems}
\label{interface_progress_water}
\end{figure}

\begin{figure}[htb]
\centering
\includegraphics[width=0.5\textwidth]{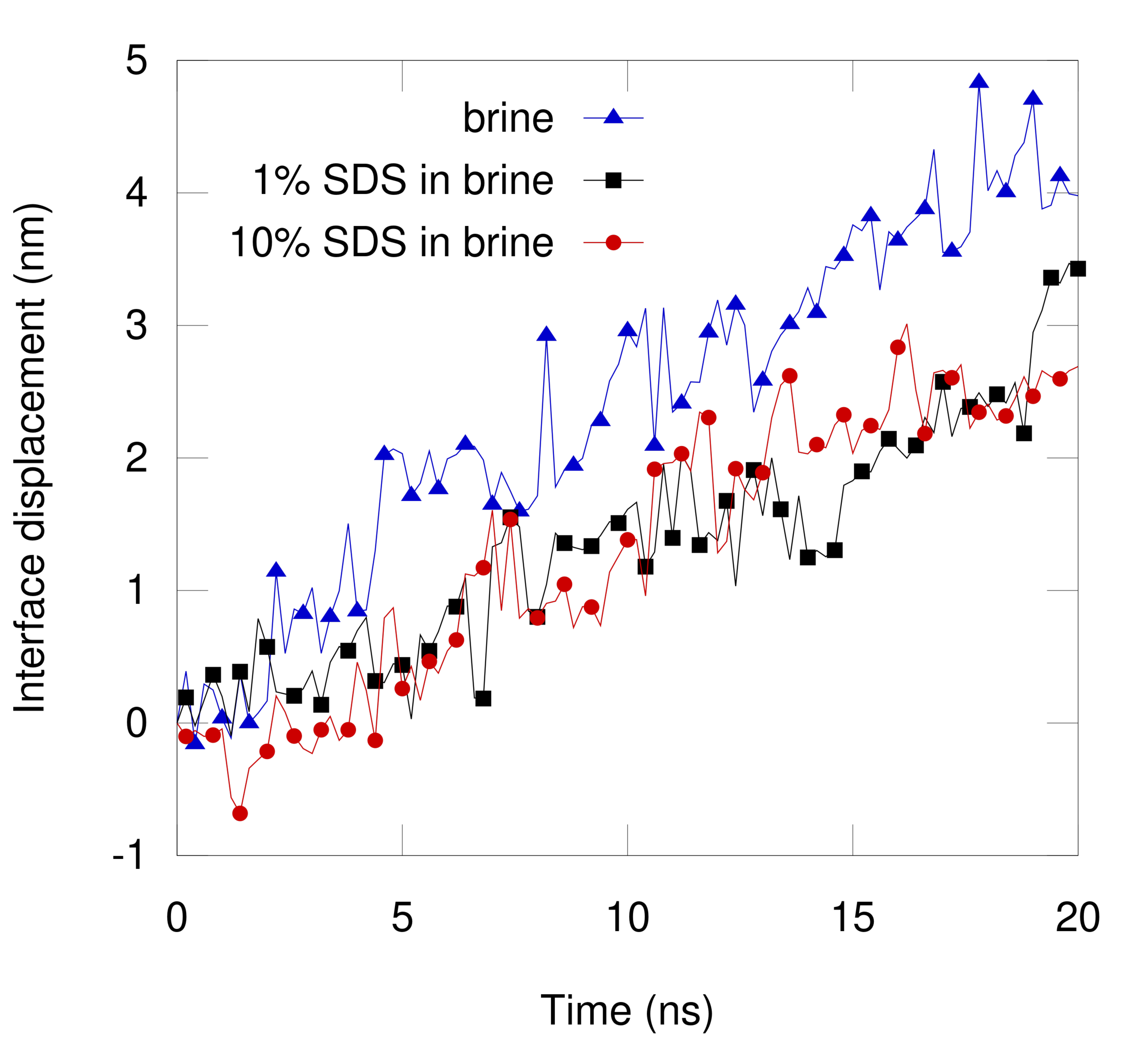}
\caption{Interface evolution in the spontaneous fluid-fluid displacement for the brine-SDS-dodecane systems}
\label{interface_progress_brine}
\end{figure}

Figures \ref{interface_progress_water} and \ref{interface_progress_brine} show the time series of the spontaneous ({\it i.e.}, driven solely by the capillary pressure) interface displacement considering, respectively, water and brine displacing fluids. 
The effect of added surfactants in different concentrations (1\% or 10\%) can also be observed in these plots.

A first glance at these figures reveals that the interface displacement is not a straightforward process, showing pronounced fluctuations (advancing and receding motions). The receding motions are due to curvature changes that precede the advancing motion. 
In other words, the capillary forces promote the displacement of the molecules in the proximity of the internal walls of the tube, leading to a deformation of the curvature, which is accompanied by an increase in interfacial tension. Therefore, to alleviate this tension, a reorganization of the molecules close to the interface occurs, recomposing the equilibrium curvature and producing a net effect of an advancing motion.
In spite of these fluctuations, the observed progress was almost linear for all cases, which is in line with the work of Chen \textit{et al} \cite{Chen2014}. 

Considering the water systems, while no significant changes could be noticed for the wettability, as it is pointed out by the contact angle analysis (see Table \ref{cap_contact_angle}), the surfactant molecules led to a lower interfacial tension. This explains the fact that the addition of surfactants resulted in a lower velocity of the interface. Following the MKT (Equation \ref{Blake1}) \cite{blake1969}, the fluid-fluid displacement is an activated process favored in one direction and hampered in the other according to the work done by the surface forces. The lower the interfacial tension, the lower the work done by the surface forces and slower will be the interface velocity in the favored direction.

Considering  the brine systems, the decrease in velocity upon adding surfactant is less pronounced and no significant differences were observed by increasing the surfactant concentration from 1\% to 10\%, since the lower surfactant concentration was sufficient to reduce the interfacial tension to nearly zero.  In order to compare the results for water and brine systems one has to keep in mind that they differ in viscosity. The higher viscosity of brine may explain the slower interface progress in all cases when compared with water.

\begin{figure}[htb]
\centering
\includegraphics[width=0.5\textwidth]{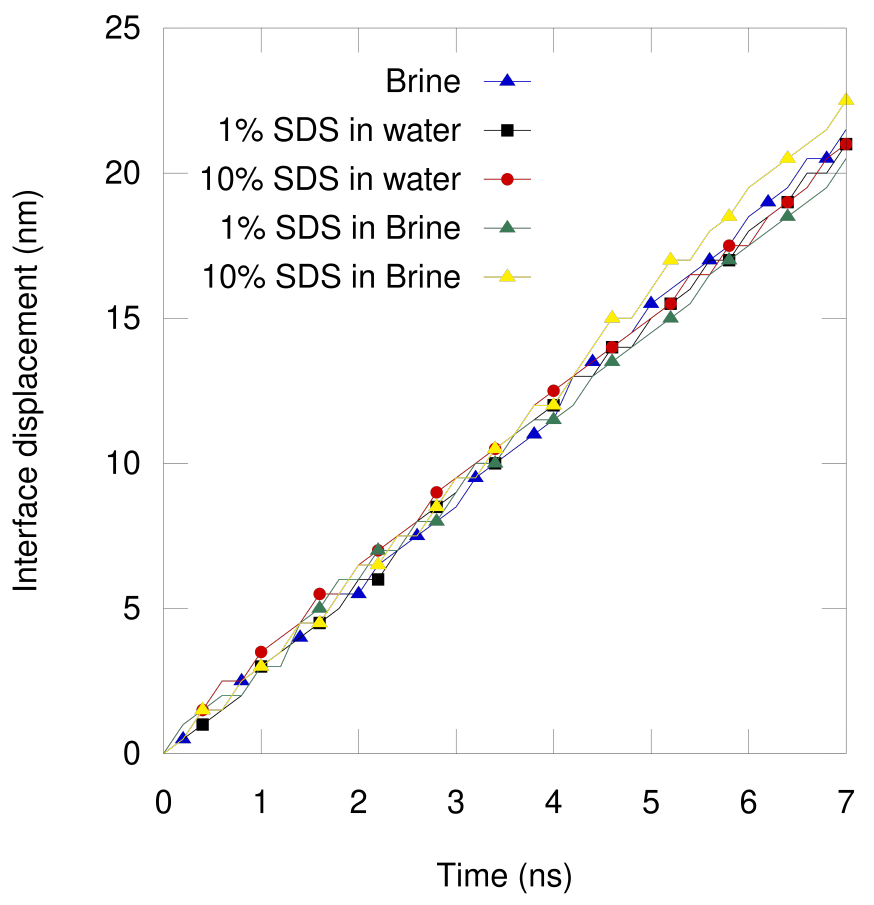}
\caption{Interface evolution in the forced fluid-fluid displacement.}
\label{forced}
\end{figure}

The forced fluid-fluid displacement simulation was carried out by introducing a virtual piston acting on the injection fluid reservoir as described in the methods Section. Upon moving, the piston  pushes the beads at the boundary of the reservoir, imposing a velocity field that is transferred by intermolecular interactions to the other particles along the vessel. At the steady state,  the beads acquire an average velocity that depends solely on the piston velocity. Figure \ref{forced} shows that the different systems presented the same interface velocity, as it was expected due to the conservation laws.


The piston velocity was set to  8 \AA /ns. All the systems containing 10 \% of SDS and the system consisting of 1 \% of SDS in brine, when subjected to the action of this piston, presented a protrusion behavior. In other words, a non-wetting behavior is developed due to the fact that the surfactant turned the interactions between water and oil molecules more favorable then the interactions between the water and wall particles. It is important to say that due to this protrusion it was impossible to fit the spherical cap to the interface. Because of that it was decided to track the position where the local density was approximately the average density of both fluids in order to construct the interface displacement plots. In the systems where the protrusion occurred, another simulations was performed with a halved piston velocity.

The microscopic contact angle, which within the MKT is equivalent to the dynamic contact angle, is uniquely defined only when thermal fluctuations of the interface are averaged out \cite{Thompson1989,  Bonn2009}. Usually, in non-equilibrium molecular dynamics simulations, one waits for a steady state to develop and extract relevant average quantities. For example, Thompson and Robbins \cite{Thompson1989} performed simulation of Couette flows of Lennard-Jones fluids and extracted the dynamic contact angle for each surface velocity by calculating the slope of the average velocity profile of the interface. In this work, it was considered that a steady state was reached when the instantaneous microscopic contact angle started to fluctuate around an average value, which was considered to be the microscopic contact angle. The calculated  values are displayed in Table \ref{cap_contact_angle} considering piston velocities of 0, 4 and 8 \AA /ns. 

Interestingly, the presence of brine or surfactant increased the time for the systems to reach the stationary state for the contact angle in both spontaneous and forced displacements. In the system of only water and dodecane, it took about 10 ns for the contact angle to fluctuate around the average, but in the other systems, this time increased to about 20 ns. It is important to keep in mind that these timescales only reflect a higher degree of complexity of the systems containing additives. In this case, quantitative results may only be reached by considering an ensemble of trajectories.

\begin{table}[htb]
\centering
\begin{threeparttable}[b]
\caption{\label{cap_contact_angle} Average microscopic contact angle in degrees for different piston velocities in \AA /ns}
\begin{tabular}{cccc}
\hline
   System & 0 & 4 & 8  \\ \hline
 Water & 76  & -\tnote{a} & 131  \\
 1\% SDS in water & 70  & -\tnote{a} & 144 \\ 
 10\% SDS in water & 76 & -\tnote{b} & -\tnote{b}  \\ 
 Brine & 70 & -\tnote{a} & 116 \\
 1\% SDS in brine & 82 & 116 & -\tnote{b} \\
 10\% SDS in brine & 73 & -\tnote{b} & -\tnote{b}  \\ \hline
\end{tabular}
   \begin{tablenotes}
     \item[a] Not simulated.
     \item[b] Not able to calculate the contact angle due to protrusion behaviour.
   \end{tablenotes}
\end{threeparttable}
\end{table}

%
%

The standard deviation of the contact angle was about 6\degree{} for all systems.
Therefore, all values calculated for the case of  spontaneous displacement are statistically indistinguishable. Upon increasing the interface velocities by moving the piston, an increase in the contact angle is observed for all systems, as expected according to the MKT. 
The effect is more pronounced for systems with a lower interfacial tension: ($i$) comparing water and brine, water has a larger increase in contact angle; ($ii$) comparing water or brine with their respective surfactant mixtures, both surfactant-containing systems evidence a larger increase in contact angle. Just note that, under a piston velocity of 4 \AA /ns, the 1\% SDS brine solution has an equivalent contact angle compared to the system without SDS calculated at 8 \AA /ns. 
%

Figure \ref{nobrine_interface} displays the last frame of the simulations involving the system consisting of 1 \% SDS in water.  The top panel shows the system without the action of the piston, reflecting that, in a low velocity regime, the capillary forces dominate the process and the injection fluid advances by wetting the surface. The bottom panel shows the system under the action of a piston moving at 8 \AA /ns velocity. The imposed flow changes the displacing mechanism and this is visually evident by comparing the two interfaces of Figures \ref{nobrine_interface}a and \ref{nobrine_interface}b. 
It is also interesting to note the behaviour of the surfactant molecules that, in the case of low velocity, are spread all over the interface, whereas in the case of an imposed flow, they concentrate in the central front, also evidencing an aggregation behaviour resembling that of reverse micelles.

%
The occurrence of the reverse micelles structures could not be detected in the simulations carried out to calculate the interfacial tension as a function of the amount of surfactant adsorbed at the interface. Those simulations were carried out with periodic boundary conditions without the presence of a wall and only the assembling of regular micelles were observed (See the movies attached at the Supplementary Material). Therefore, it could not be predicted that reverse micelles could be formed in the conditions such as those of the simulations involving the silica channel. 
%

Figure \ref{brine_interface} shows the last frame of the simulation of the system containing 1\% SDS in brine. A similar dependency of the curvature with the velocity is observed. Without the action of the piston, the capillary forces dominate and the system behaves as water wetting.
However, one clear difference between the brine and water systems is the number of surfactant molecules at the interface that is remarkably larger in the brine system. This is in line with the experimental observations that indicate a lower interfacial tension of brine-SDS-oil systems compared to water-SDS-oil systems. Due to the more pronounced decrease in interfacial tension for the brine-SDS-oil systems, one can observe the similar curvature change that happens in water-SDS-oil systems, but in a lower velocity regime.

Moreover, the brine systems exhibited a larger degree of  aggregation of surfactant molecules at the interface. Without the action of the piston (Figure \ref{brine_interface}a), most of the surfactant molecules assemble as a bilayer structure instead of monolayer as was the case for the corresponding water system (Figure \ref{nobrine_interface}a). Under the action of the piston (Figure \ref{brine_interface}b), the behavior was similar to the one observed in the corresponding water system (Figure \ref{nobrine_interface}b), but with a much larger number of surfactant molecules at the interface. It is interesting to note that, in both cases (water and brine), it is evident that the surfactant molecules previously interacting with the internal wall of the vessel in the absence of the applied velocity migrate to the innermost region of the tube upon switching on the piston. For the brine case, the squeezing of the bilayer upon switching on the piston led to a change in aggregation behaviour, forming structures that resemble reverse micelles. 
Increasing the piston velocity to 8\AA /ns induced the leakage of the surfactant molecules through the interface to the bulk of the oil. This process occurred by forming a reverse vesicle that protruded into the oil phase (Figure \ref{brine_interface}c).
This behaviour has been reported in the context of microfluidic devices designed to the production of such structures \cite{SHAH2008}.

The systems containing 10 \% of surfactants presented the formation of bilayers at the interface, that was only possibly due to the inclusion of the cation beads between the layers. The interfacial bilayer led to the formation of a vesicle in the presence of flow. The surfactants in the bulk also form some bilayer domains which aggregated when the virtual piston was acting. Multiple movies concerning the different analyzed trajectories are made available as Supplementary Material and they show all the features discussed above 

Concerning the oil displacement, all scenarios clearly stated that the decrease of the interfacial tension coupled with high flow velocities will lead to a decrease in the oil recovery. The high contact angles in those cases are related with more dodecane molecules trapped in the region between the displacing fluid front and the channel walls (See Figures \ref{nobrine_interface}, \ref{brine_interface} and the movies attached in the Supplementary Material). Since the surfactant main effect is to decrease the interfacial tension of water-oil and brine-oil systems, those results are in line with the observations of Chen \textit{et al.} \cite{Chen2012} that favorable interactions between the fluids lead to poor oil recovery. 

As Chen \textit{et al.} \cite{Chen2012} also put, this effect may be avoided when low velocity regimes are employed. This is also true in scenarios involving surfactants as the simulations performed in this work showed no changes in the spontaneous fluid displacement with the increase of the surfactant amount. If, however, high flow regimes are needed, one would probably have to add an additive to increase the displacing fluid viscosity, such as polymers, to avoid the poor recovery, since this additives will increase the sweep efficiency.

\subsection{Polymer effects}

The primary effect of polymers is to increase the sweep efficiency of the displacing fluid by increasing its viscosity. However, it is still an open question how to observe such effects considering the time and size scales currently achievable  with molecular simulations. 
Due to the inherent challenges of accurately and precisely quantifying such effects, the results concerning polymer effects are described on a qualitative basis. 

\begin{figure}[H]
\centering
\includegraphics[width=0.5\textwidth]{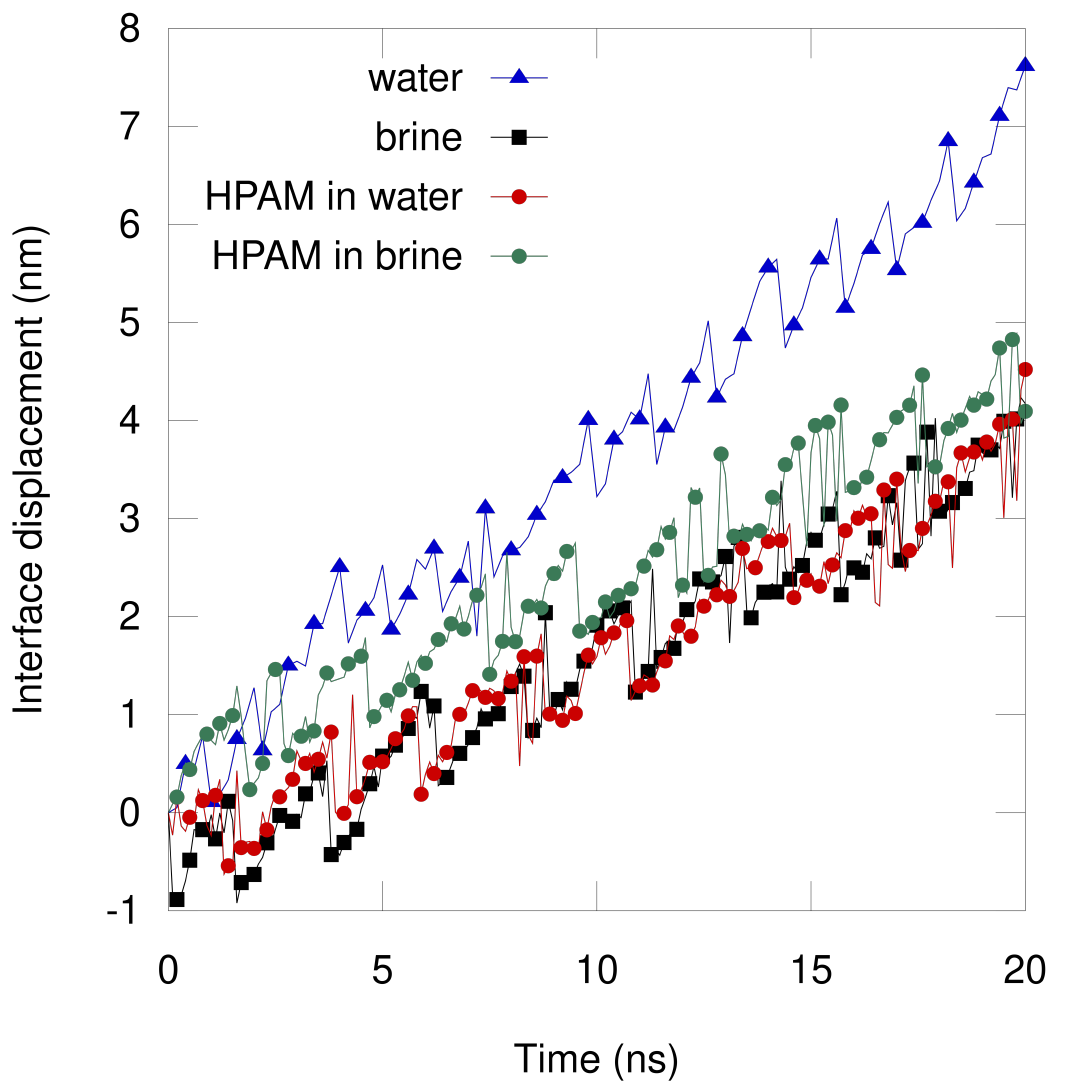}
\caption{Interface advance in the spontaneous fluid-fluid displacement with the presence of HPAM.}
\label{pol-interface-advance}
\end{figure}

Figure \ref{pol-interface-advance} shows that the presence of HPAM in water slows down the interface advance in the spontaneous displacement scenario. However, this may not be related to a viscosity increase. As Figures \ref{polymer-water} and \ref{channel-entrance} display, another explanation for this effect may be the adsorption of polymer molecules at the reservoir wall at the channel entrance. This was not observed in the simulations with surfactants even when they were randomly inserted in the water reservoir.

The polymer adsorption at the reservoir walls are suppressed in the presence of brine as Figures \ref{polymer-brine} and \ref{channel-entrance} show. This is also suggested by the fact that the interface velocity is very similar in the cases of brine and HPAM in brine, as displayed by the plots in Figure \ref{pol-interface-advance}. These results show that the HPAM polymer does not affect the balance between the fluid-wall and fluid-fluid interactions as do the surfactant species. 

This happened because the polymer experience more favorable interactions with the beads that constitute the brine than with the pure water beads. This increase in the attraction is probably sufficient to overcome the interaction of the polymer with the wall and avoiding the adsorption. This results are interesting because they show that clogging of capillary tubes would be avoided in the presence of counterions in the polymer solution.

Also of note is the fact that the interface displacement was similar for the system of HPAM in pure water and the systems containing brine. Probably, the adsorption of the polymer slows down the water dynamics in a way that made it comparable to the slower dynamics of the more viscous brine solution.

Concerning the forced fluid displacement simulations, the imposed flows were sufficient to overcome the adsorption of the polymer molecules (See Figures \ref{polymer-water}b and \ref{polymer-brine}b). As expected, the HPAM molecules do not affect the dynamics of the interface as they barely reach the interface region. Furthermore,  no significant differences between the systems containing water and brine could be noticed at the interface.

\FloatBarrier

\section{Conclusion}

In this work, an MDPD model was developed to study the effects of additives on the water-oil and brine-oil displacement occurring inside capillary tubes of nanometric dimensions. The model was derived in order to reproduce both experimental and atomistic simulation results. The model describes
the decrease of the interfacial tension with the increase of the surfactant amount and the preferential interactions between the polymer units and the different species in the environment. 

In relation to the fluid-fluid displacement, no differences were observed for the microscopic angle when the spontaneous displacement phenomena was studied for the different cases involving water, brine and different amounts of surfactants, but the presence of surfactants and brine led to a slower interface evolution. On the other hand, the simulations of the high velocity regimes indicate that even this chemically heterogeneous system seems to obey the MKT, at least qualitatively. The combination of such regimes and high surfactant concentrations changed the displacement mode leaving a n-dodecane rich region between the capillary walls and the fluid-fluid interface.

Those findings suggest that, when oil is to be recovered from filled nanocapillaries in  high velocity regimes, surfactants could not be used alone or there will be oil molecules trapped at the region between the advancing front and the solid walls. The simulations presented in this work also suggest that the use of HPAM should be combined with a displacing fluid containing counterions, such as brine, in order to avoid adsorption at the reservoir walls.

The presented simulations in the highest velocity regimes pointed out that these systems are able to drive the assembly of micelles and vesicles and the present methodology may be used to help the understanding of the different phenomena related to this processes. In order to do so, the parametrization strategy should probably include the description of structural characteristics of micelles obtained by means of classical molecular dynamics or any microscopical technique.

\section{Acknowledgements}

GCQS would like to thank CNPq for providing a fellowship. This research was also carried out with the support of the Rensselaer Polytechnic Institute (RPI). FAPERJ and CAPES (code 001) are also acknowledged. 

\section{Competing Interests}
The authors declare no competing financial interests.




  
\newpage  
  
\begin{figure*}[htb]
\center
\subfloat[Full simulation box]{
\includegraphics[scale=0.2]{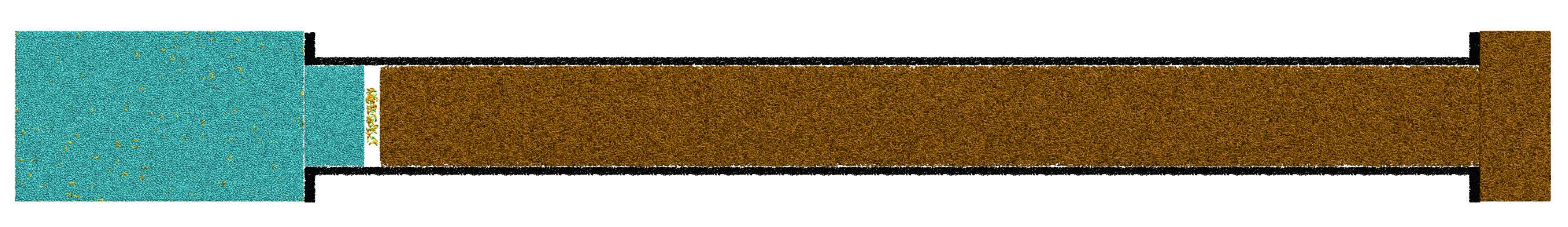}}
\end{figure*}

\begin{figure*}\ContinuedFloat
\center
\subfloat[Zoom at the interface zone of the simulation box]{
\includegraphics[scale=0.25]{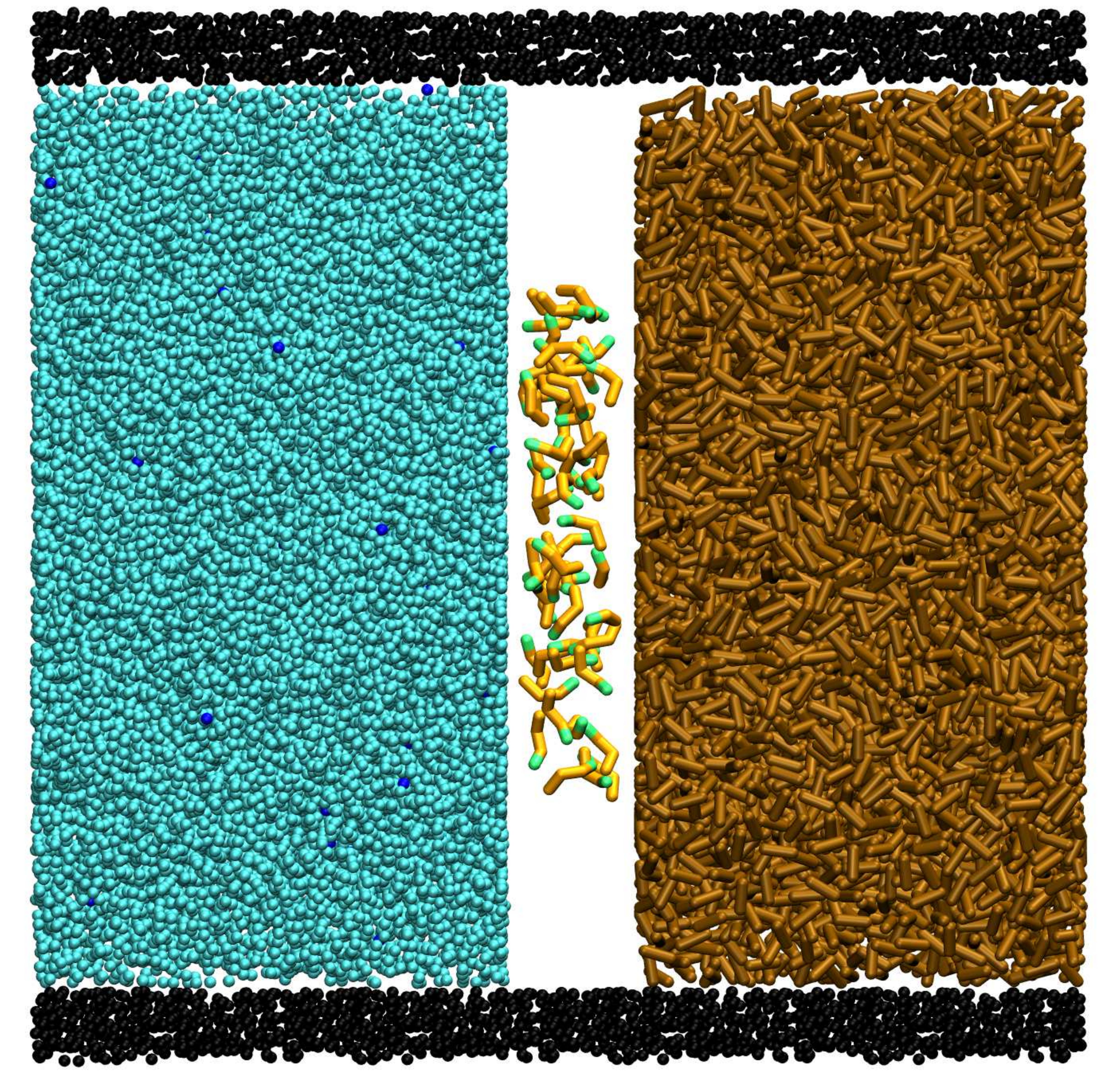}}
\end{figure*}

\begin{figure*}\ContinuedFloat
\center
\subfloat[Tilted view of the capillary walls]{
\includegraphics[scale=0.25]{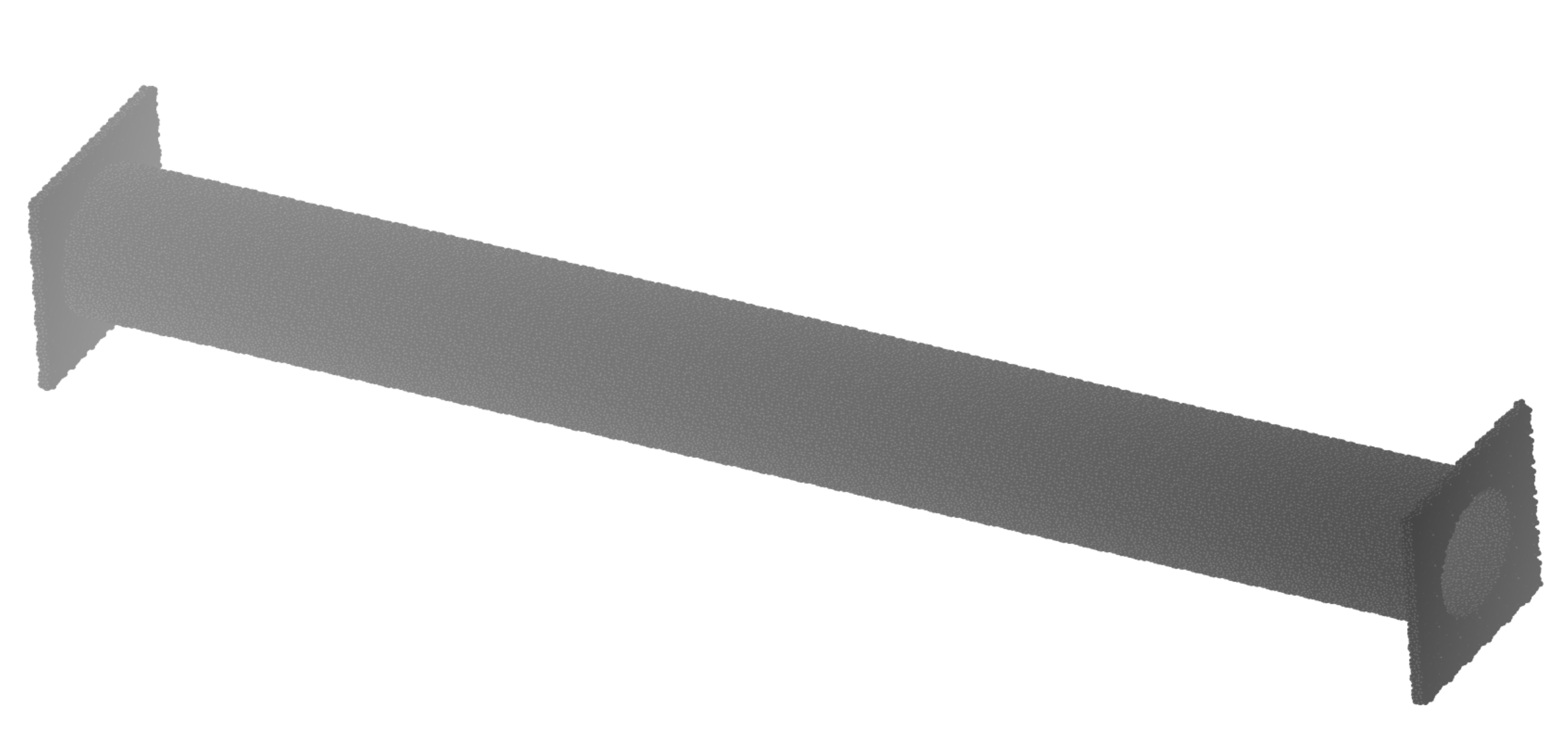}}
\end{figure*}

\begin{figure*}\ContinuedFloat
\center
\subfloat[Tilted view of the capillary without the cylinder wall]{
\includegraphics[scale=0.25]{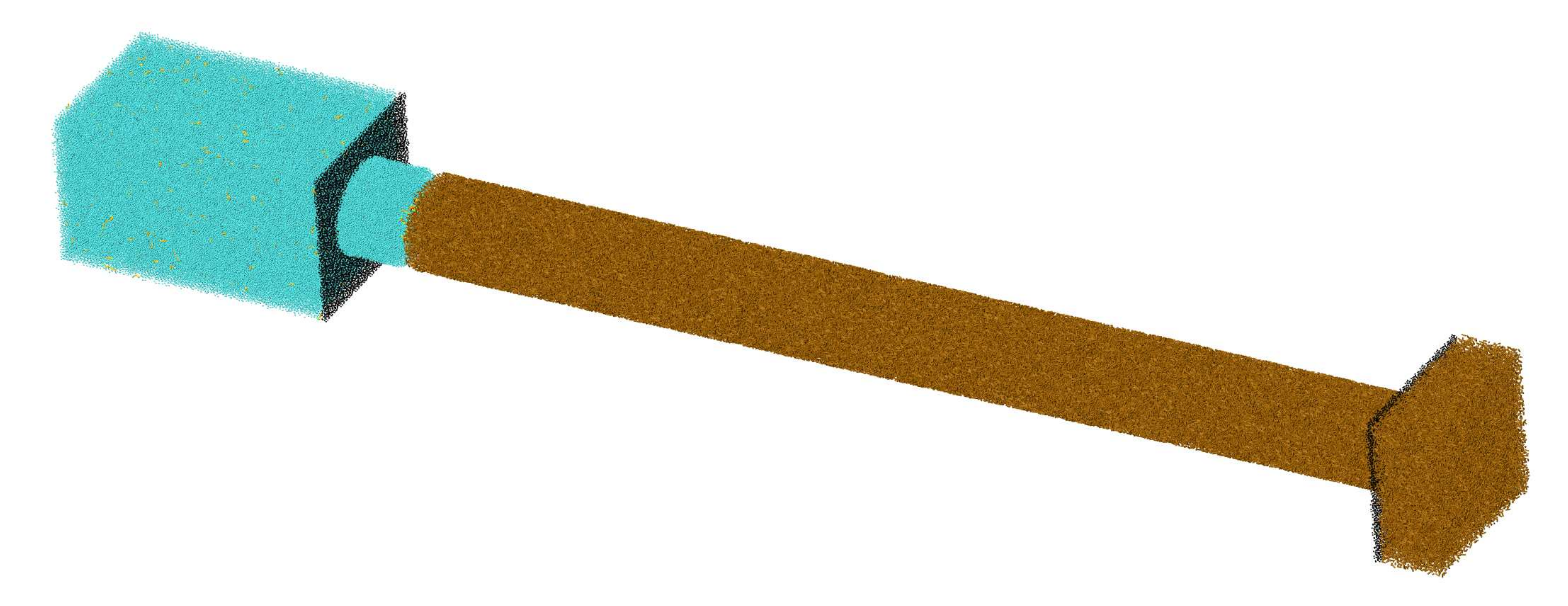}}
\setcounter{figure}{7}
\caption{Example of an initial setup of the simulation}
\label{system}
\end{figure*}

\begin{figure*}[htb]
\center
\subfloat[No virtual piston]{
\includegraphics[scale=0.15]{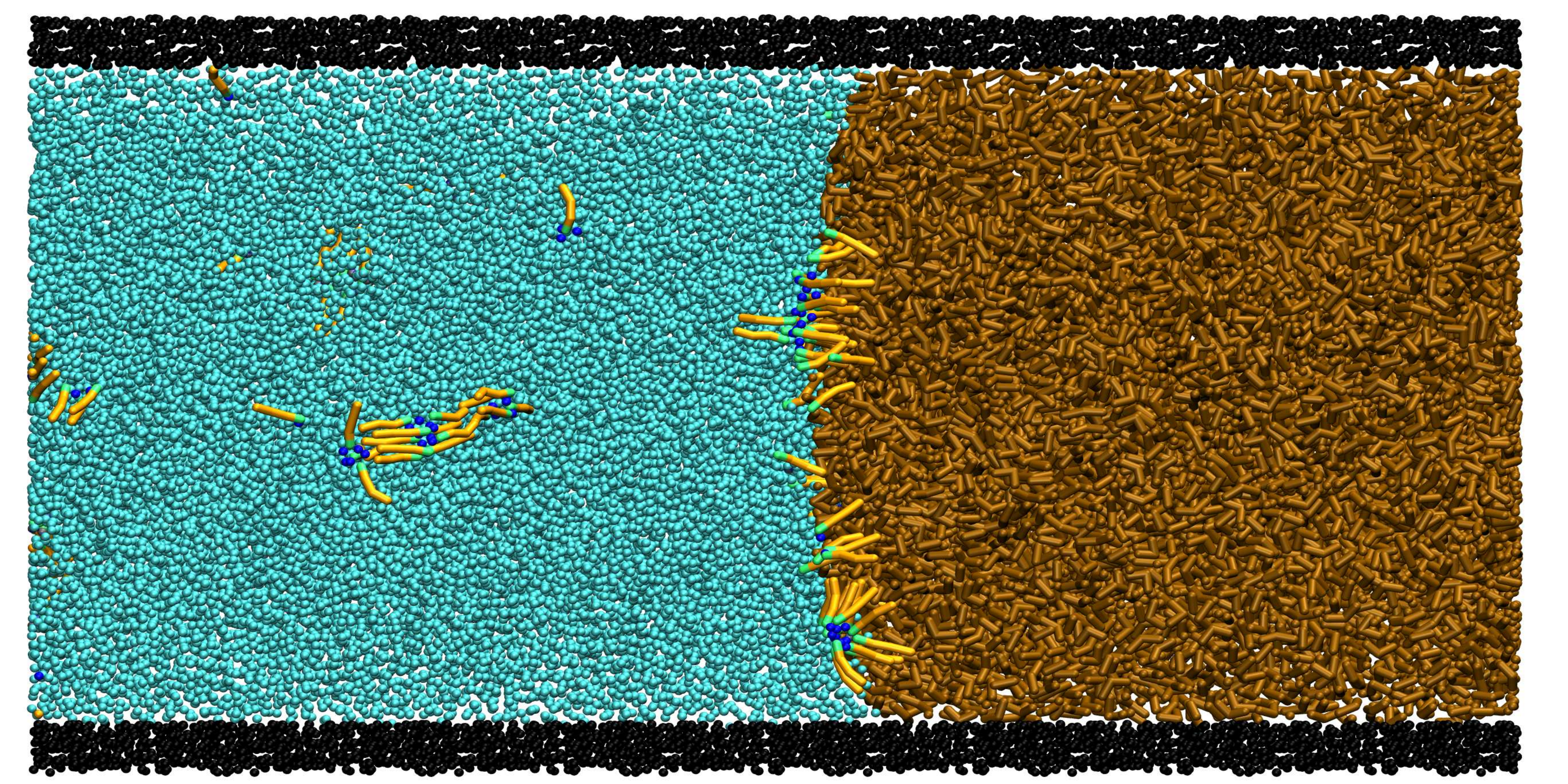}}

\subfloat[Piston velocity of 8 \AA /ns]{\includegraphics[scale=0.15]{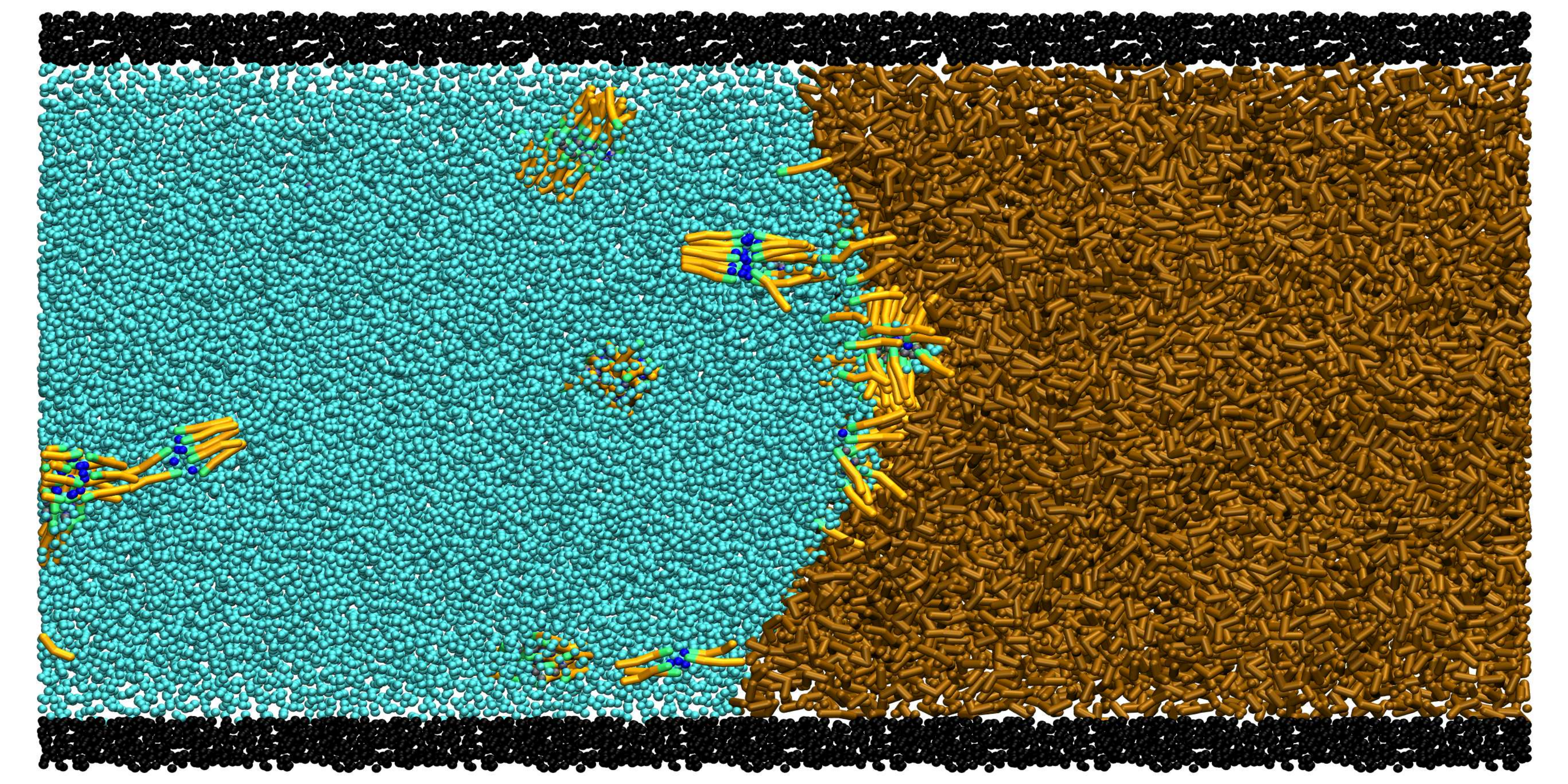}}
\setcounter{figure}{8}
\caption{Simulation snapshots of the interface in the systems consisting of 1 \% SDS in water}
\label{nobrine_interface}
\end{figure*}

\begin{figure*}[htb]
\center
\subfloat[No virtual piston]{
\includegraphics[scale=0.15]{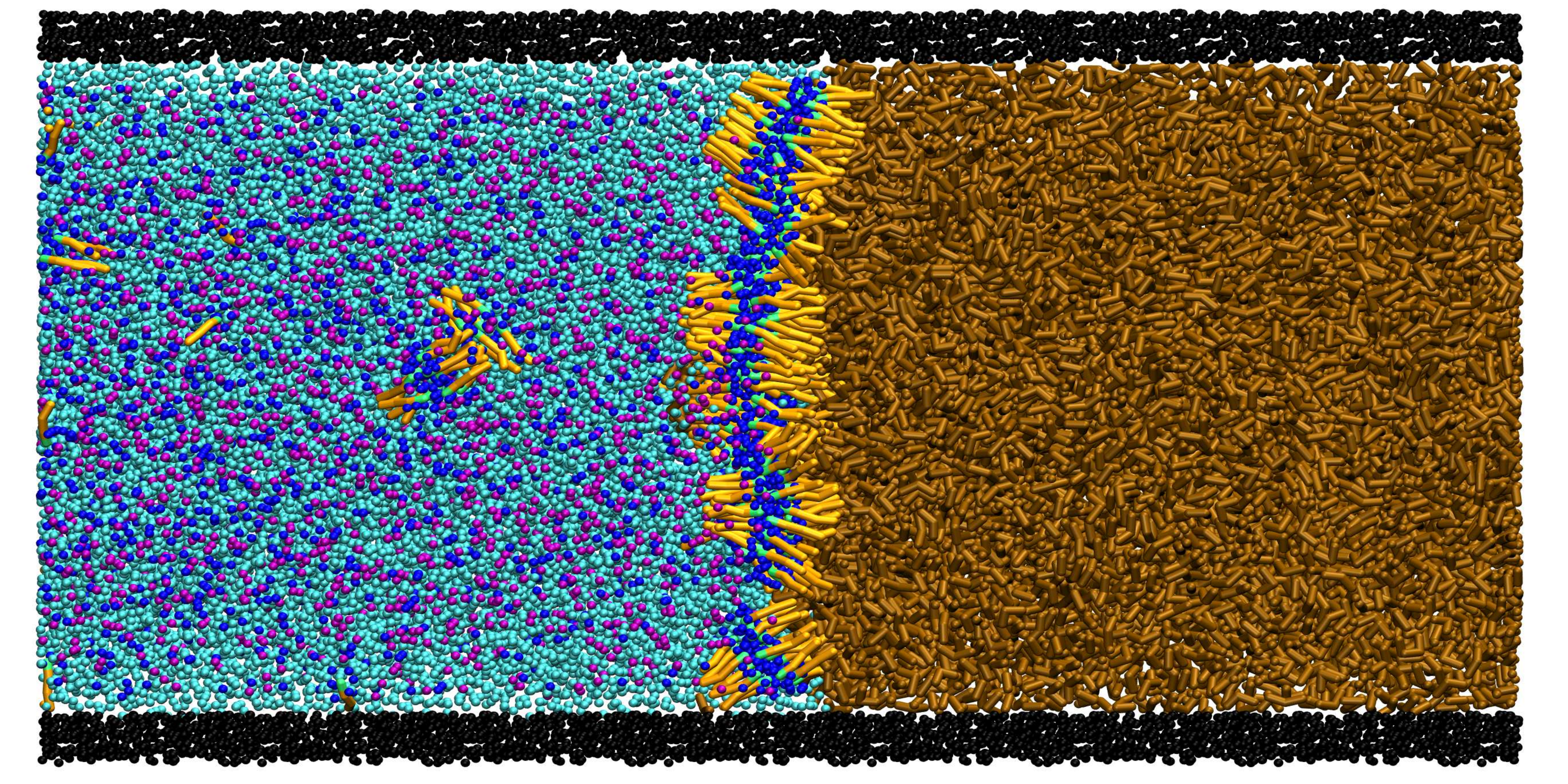}}
\end{figure*}
\begin{figure*}
\center
\ContinuedFloat

\subfloat[Piston velocity of 4 \AA / s]{\includegraphics[scale=0.15]{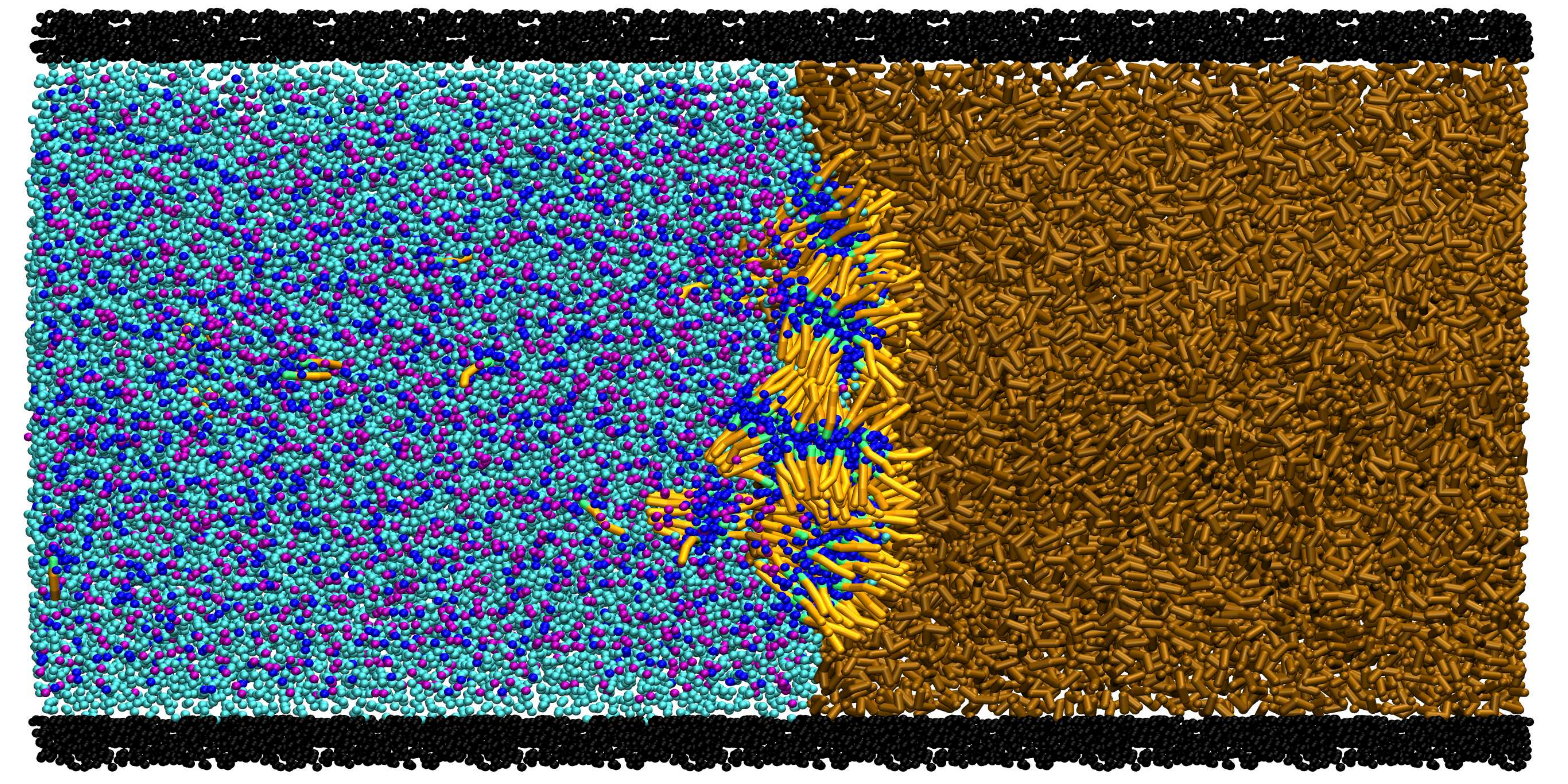}}

\end{figure*}

\begin{figure*}
\center
\ContinuedFloat

\subfloat[Piston velocity of 8 \AA /ns]{
\includegraphics[scale=0.15]{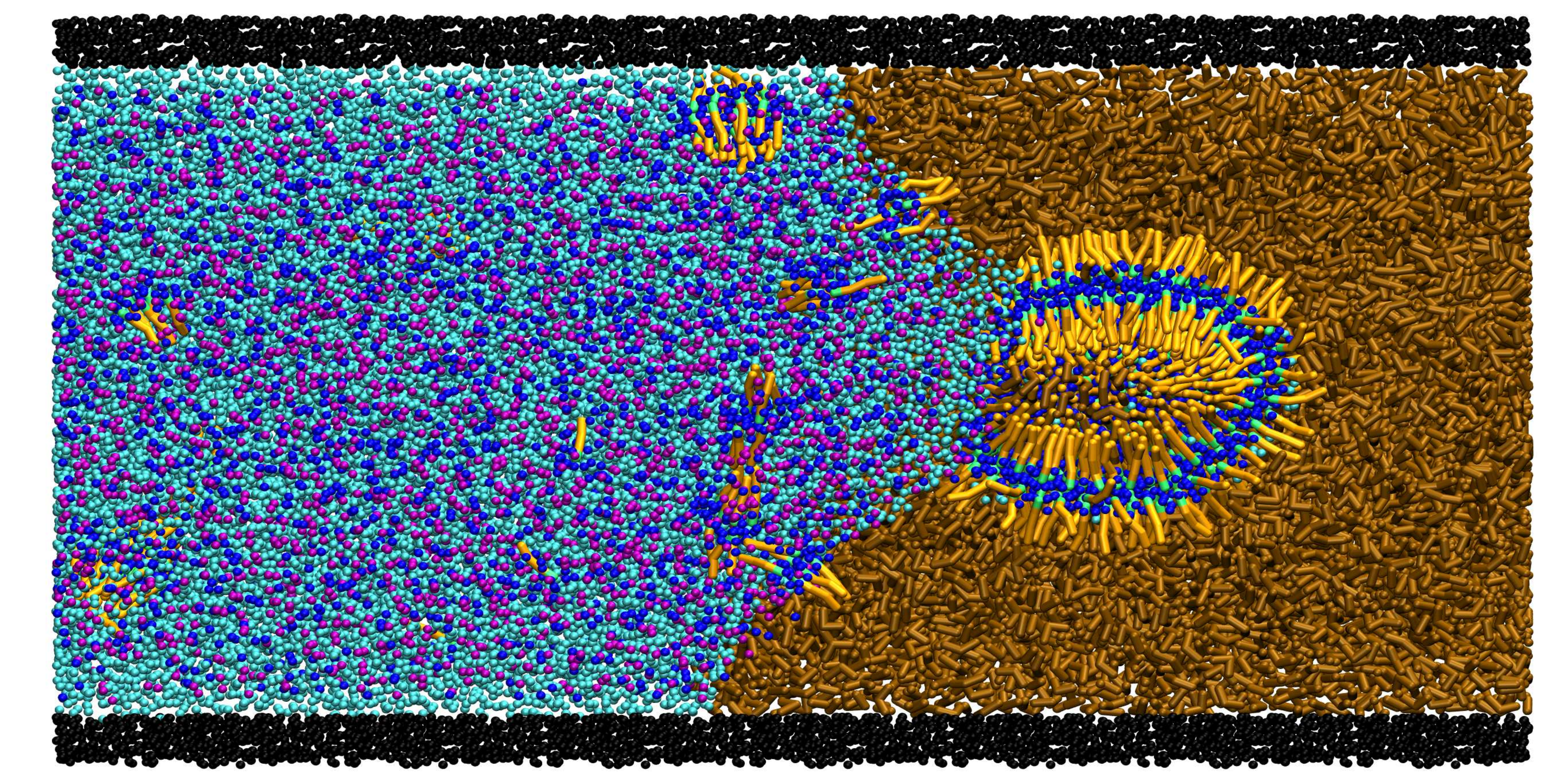}}
\setcounter{figure}{9}
\caption{Simulation snapshots of the interface in the systems consisting of 1 \% SDS in brine}
\label{brine_interface}
\end{figure*}

\begin{figure*}[htb]
\center
\subfloat[Spontaneous fluid displacement]{
\includegraphics[scale=0.22]{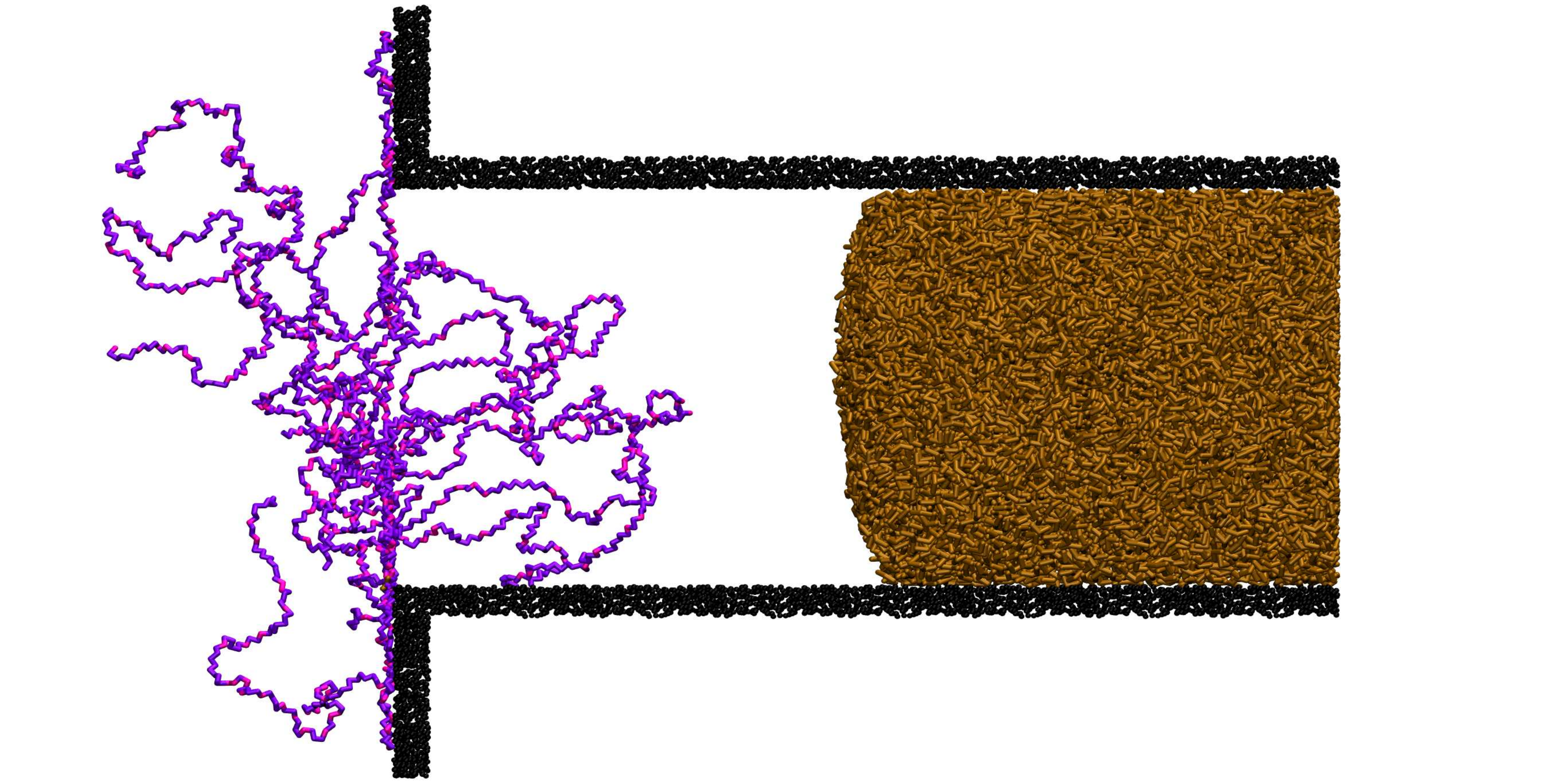}}
\center
\subfloat[Forced displacement]{
\includegraphics[scale=0.2]{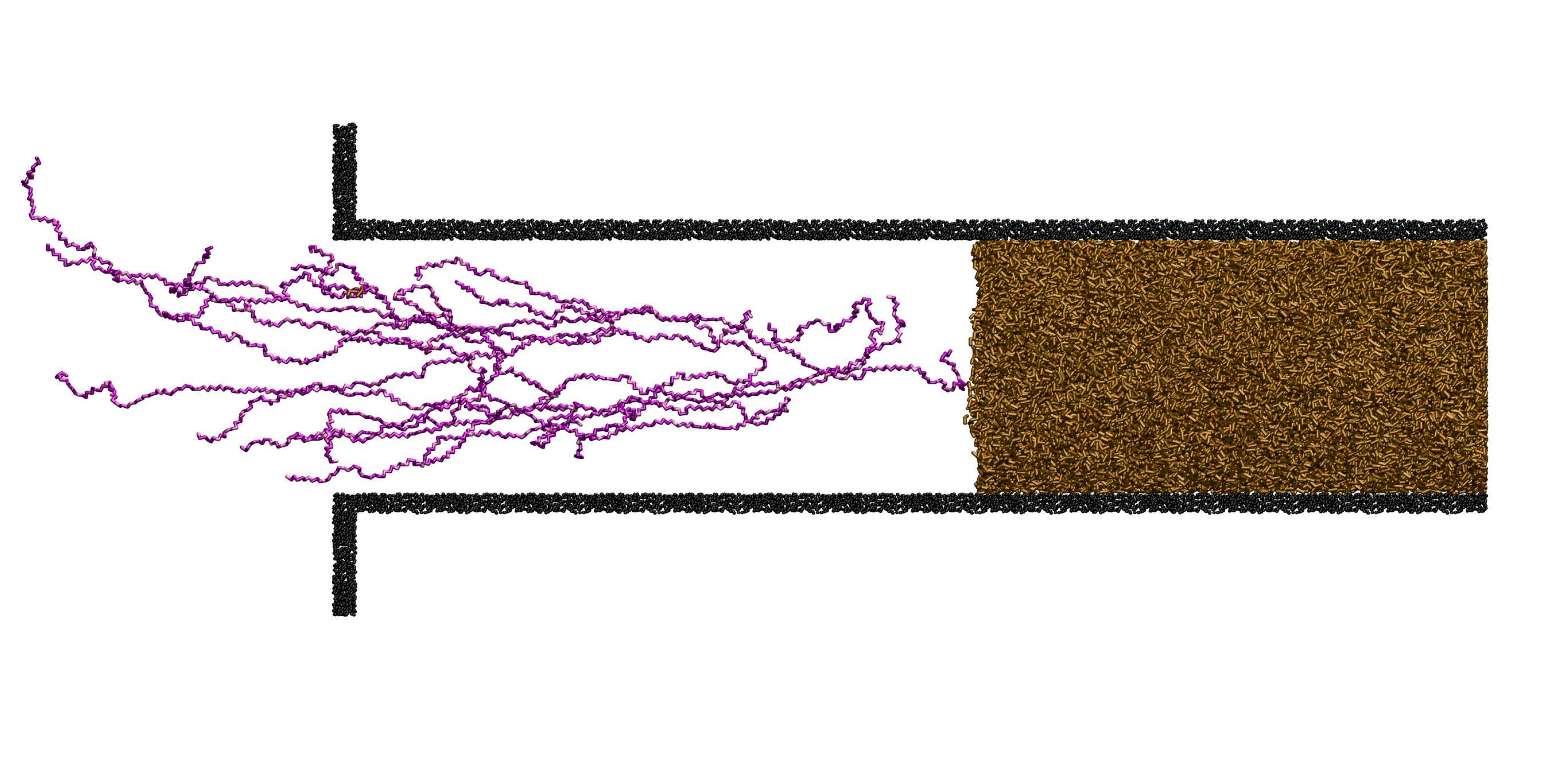}}
\setcounter{figure}{10}
\caption{Simulation snapshots of the fluid-fluid displacement with HPAM in water (Water beads are omitted)}
\label{polymer-water}
\end{figure*}

\begin{figure*}[htb]
\center
\subfloat[Spontaneous fluid displacement]{\includegraphics[scale=0.22]{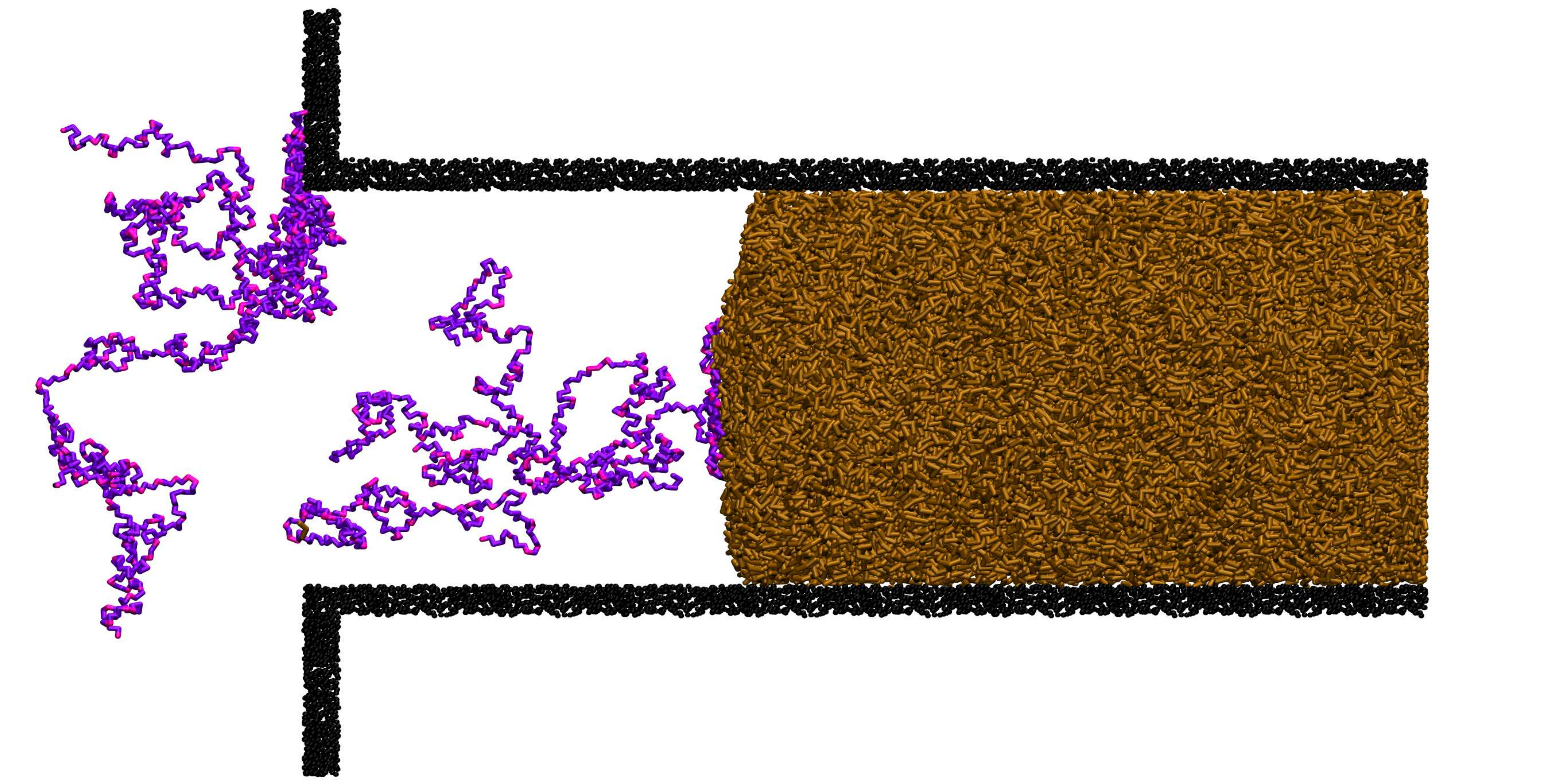}}
\center
\subfloat[Forced displacement]{\includegraphics[scale=0.2]{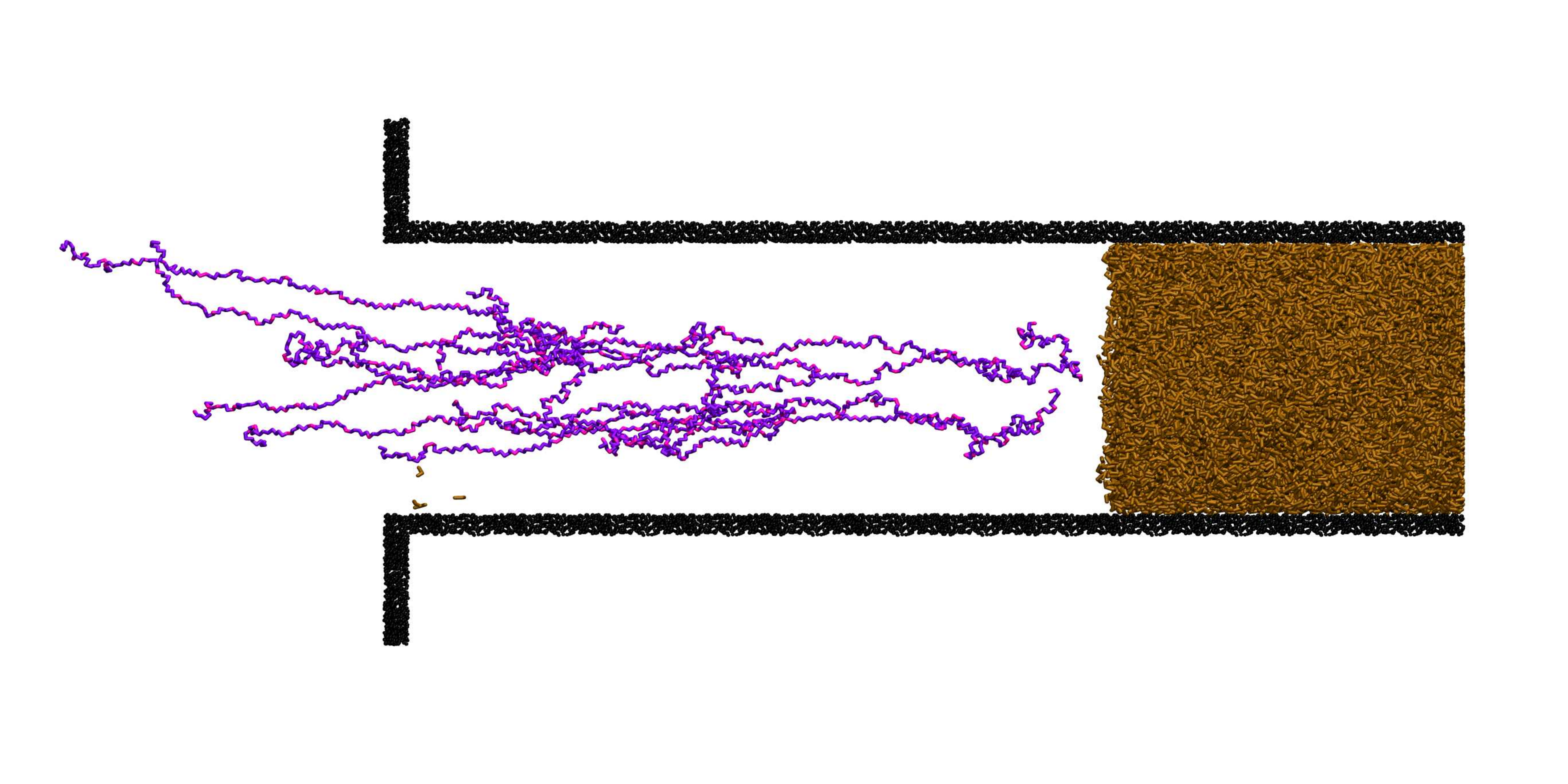}}
\setcounter{figure}{11}
\caption{Simulation snapshots of the fluid-fluid displacement with HPAM in brine (Water and ion beads are omitted)}
\label{polymer-brine}
\end{figure*}
\begin{figure*}[htb]
\center
\subfloat[HPAM in water]{
\includegraphics[scale=0.2]{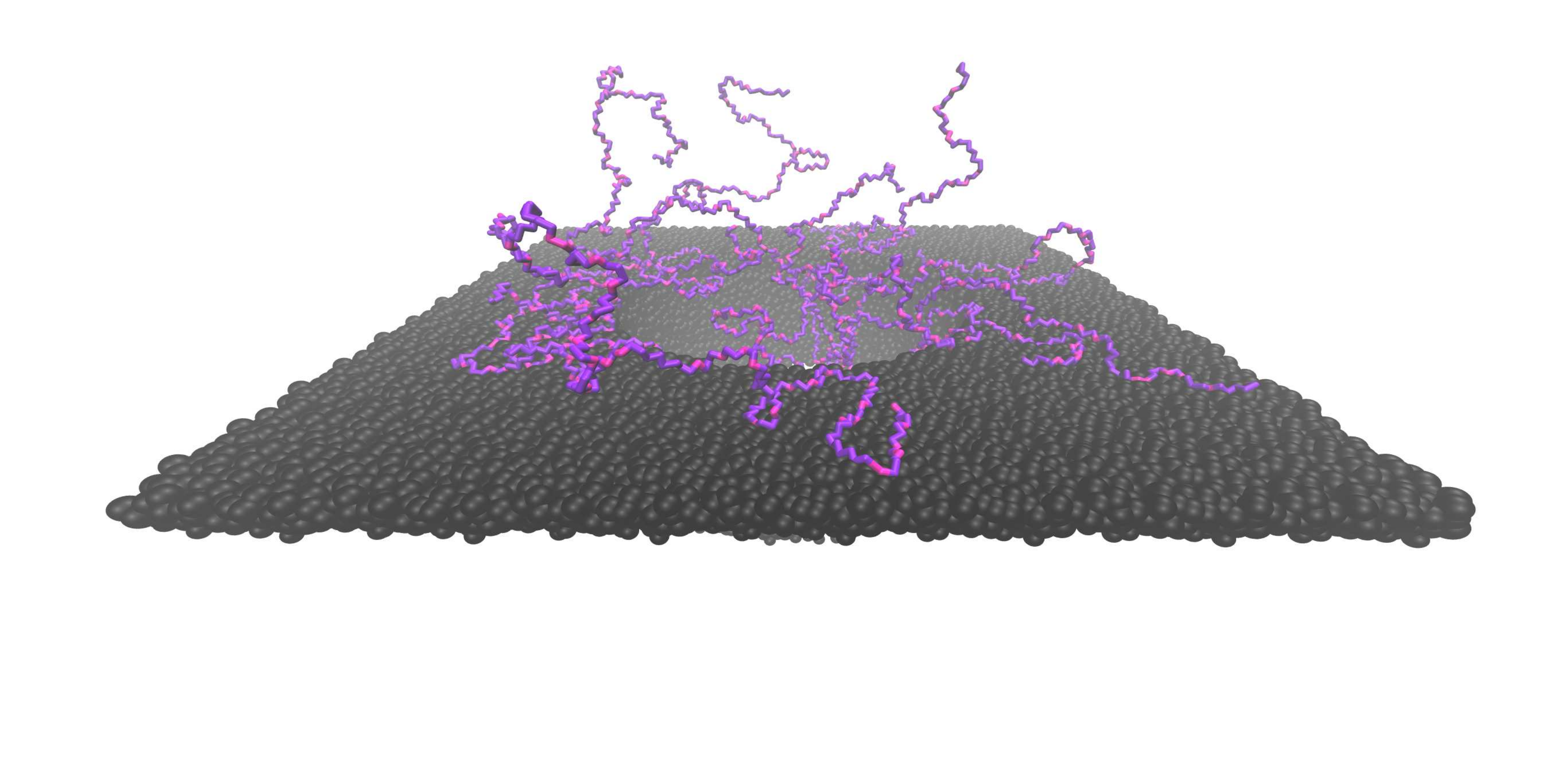}}
\center
\subfloat[HPAM in brine]{
\includegraphics[scale=0.2]{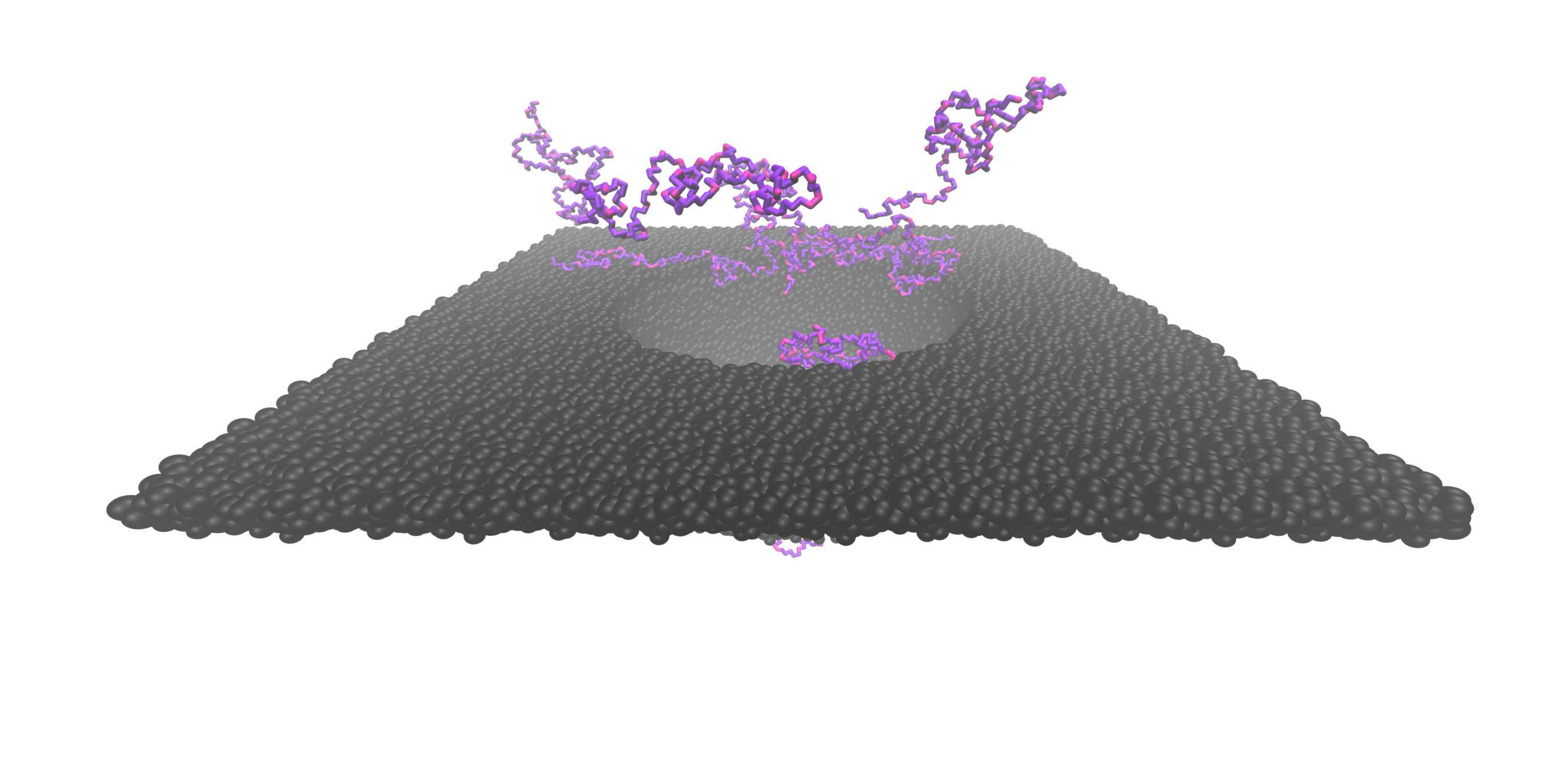}}
\setcounter{figure}{12}
\caption{Simulation snapshots of spontaneous fluid-fluid displacement at the capillary entrance (water and ions beads are omitted)}
\label{channel-entrance}
\end{figure*}
%






\newcommand{\beginsupplement}{%
        \setcounter{table}{0}
        \renewcommand{\thetable}{S\arabic{table}}%
        \setcounter{figure}{0}
        \renewcommand{\thefigure}{S\arabic{figure}}%
     }

\newpage
\onecolumn

\section{Supplementary Material}
\beginsupplement

\subsection{Fitting of the MDPD model}

\begin{figure}[h]
\includegraphics[width=\textwidth]{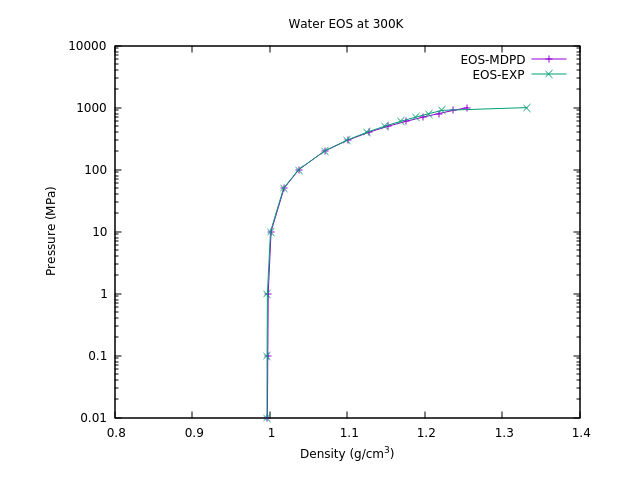}
  \caption{Water PV diagram at 300 K EOS: comparison of the MDPD model and the experimental data reported by Wagner and Pruss (2002) \cite{Wagner2002}. }
  \label{pvdiag}
\end{figure}

This supplementary material displays how the developed MDPD model reproduces the measured data used in the fit. More specifically, Figure \ref{pvdiag} and Table \ref{comparison} compare the calculated properties obtained with the MDPD model and the corresponding reference values used in the fitting procedure. These values were obtained either from experimental values previously reported in the literature or from the all-atom MD simulations performed in this work.

\begin{table}[htb]
\centering
\begin{threeparttable}[b]
\setlength{\tabcolsep}{4pt}
\caption{\label{comparison} Thermodynamic, dynamic and geometric properties used for fitting the MDPD model developed in this work. Calculated values are reported along with the reference data used as target in the calibration procedure. The reference values correspond to either experimental data previously reported in the literature or data obtained from classical molecular dynamics simulations performed in this work.}
\begin{tabular}{@{}ccc@{}}
\toprule
                                                  & MDPD & Reference \\ \midrule
$\rho_{water}$ (g/cm\textsuperscript{3})                           & 0.996      & 0.997\tnote{a}                        \\
$\rho_{dodecane}$ (g/cm\textsuperscript{3})                        & 0.750      & 0.745\tnote{a}                        \\
$\eta_{water}$ (cP)                         & 0,905      & 0.854\tnote{a}                      \\
$\eta_{dodecane}$ (cP)                            & 1,400      & 1.330\tnote{a}                        \\
$\Gamma_{water-dodecane}$ (mN/m)                  & 54,00      & 52,55\tnote{a}                       \\
$\Gamma_{water-SDS-dodecane near the CMC}$ (mN/m) & 7,4        & 9,1-10,7\tnote{b}                     \\
$\theta\degree_{\phantom{a} (water-silica-vapor)}$ (\degree)         & 20,5       & 21,2 \tnote{c}                        \\
$\theta\degree_{\phantom{a} (dodecane-silica-vapor)}$ (\degree)      & 33,0       & 28,0 \tnote{c}                         \\
$\theta^0_{\phantom{a} (water-silica-dodecane)}$ (\degree)      & 51,9       & 65,0\tnote{c} / 52,0\tnote{d}                  \\
$Rg_{\phantom{a}
(PA-dodecane)}$ (\AA)\tnote{e}                          & 21,0       & 24,7\tnote{c}                       \\
$Rg_{\phantom{a}
(PA-water)}$ (\AA)\tnote{e}                            & 17,6       & 16,4\tnote{c}                      \\
$Rg_{\phantom{a}(PN-dodecane)}$ (\AA)\tnote{e}                         & 7,8        & 8,6\tnote{c}                       
 \\
$Rg_{\phantom{a}
(PN-water)}$ (\AA)\tnote{e}                       & 11,7       & 11,6\tnote{c}                         \\
$Rg_{\phantom{a}
(PAPN-water)}$ (\AA)\tnote{e}                           & 14,1       & 14,0\tnote{c}                         \\
$Rg_{\phantom{a}
(PA-brine)}$ (\AA)\tnote{e}                             & 11,5       & 7,1\tnote{c}                          \\
$Rg_{\phantom{a}
(PN-brine)}$ (\AA)\tnote{e}                             & 10,5       & 10,1\tnote{c}   
\\
\bottomrule
\end{tabular}
\begin{tablenotes}
     \footnotesize{\item[a] Experimental data from NIST Chemsitry Webbook\cite{nistsite}.
     \item[b] Experimental data from Zeppieri \textit{et al.} (2001)\cite{Zeppieri2001}.
     \item[c] Value obtained with the MD simulations performed in this work.
     \item[d] Experimental data from Bi \textit{et al.} (2004 )\cite{BI2004}.
     \item[e] The subscripts denote the type of polymer in the metioned solvent. PN stands for the non-hydrolized polymer, PA for the fully hydrolized polymer and PAPN the partially hydrolized polymer discussed in the main text.}
   \end{tablenotes}
\end{threeparttable}
\end{table}

\twocolumn
\bibliographystyle{elsarticle-num}
\bibliography{main}


\end{document}